\documentclass[article,12pt]{elsarticle}
\usepackage{lineno,hyperref}
\modulolinenumbers[5]
\usepackage{lineno}
\modulolinenumbers[5]
\usepackage{amssymb}
\usepackage{hyperref}
\usepackage{graphicx}
\usepackage{color,soul}
\usepackage{grffile}
\usepackage{booktabs}
\usepackage{times}
\usepackage{subcaption}
\usepackage{amsthm}
\usepackage{ulem}
\usepackage[titletoc]{appendix}
\usepackage{titletoc}
\usepackage{mathtools}
\usepackage{url}
\usepackage{tikz}
\usetikzlibrary{calc}
\usepackage{multirow}
\usetikzlibrary{patterns}
\usepackage[margin=1.0in]{geometry}
\usepackage{euscript}
\usepackage{wrapfig}
\usepackage[autostyle]{csquotes}
\theoremstyle{definition}

\usetikzlibrary{matrix, shapes,arrows,positioning,chains}

\begin{document}

	\begin{frontmatter}
		
		\title{On impact of oxygen distribution on tumor necrotic region: A Multiphase Model} 	
		\author[ad]{Gopinath Sadhu\corref{mycorrespondingauthor}}
		\cortext[mycorrespondingauthor]{Corresponding author}
		\ead{gsadhu@iitg.ac.in}
		\author[f]{K S Yadav}
		\author[ab,abd]{Siddhartha Sankar Ghosh}
		\author[ad]{D C Dalal}		
		\address[ad]{Department of Mathematics, Indian Institute of Technology Guwahati, Guwahati 781039, India}
		\address[f]{Faculty of Engineering and Technology, Siksha `O' Anusandhan (Deemed to be University), Bhubaneswar 751030, India}				
		\address[ab]{Centre for Nanotechnology, Indian Institute of Technology Guwahati, Guwahati 781039, India}
		\address[abd]{Department of Biosciences and Bioengineering, Indian Institute of Technology Guwahati, Guwahati 781039, India}

\begin{abstract}
	{\bfseries Background and Objective:} In an in-vivo situation, the tissue near the blood vessels is
	rich in oxygen supply compared to the one far from blood vessels. Hence, non-uniform oxygen distribution is observed in biological tissues. Our objective is to explore the influence of non-uniform oxygen supply in the development of necrotic core and, also to examine the effects of necrotic core on tumor growth.\\ 
	{\bfseries Methods:} The research is processed through a mathematical approach based on the multiphase mathematical model. To simulate the model, a finite difference numerical method based on the “Semi-Implicit Method for Pressure-Linked Equations” (SIMPLE) algorithm is adopted.\\
	{\bfseries Results:} The necrotic core starts to form at the boundary of the tumor with lower oxygen concentration from the initial time. Investigations reveal that the position of the necrotic core varies depending on the oxygen supply through the tumor boundary. The results predict asymmetrical tumor growth under unequal oxygen supply at tumor boundaries. Also, it is hinted that a tumor with a larger size of necrotic core grows slowly as compared to a tumor containing a smaller size of necrotic core.\\
	{\bfseries Conclusions:} The formulated model has the potential to cast the situation of the tumor growth in an in-vivo and in-vitro situations. This study provides an idea about the location and shape of the necrotic core and the impact of the necrotic core on tumor growth. This information will be beneficial to the clinicians and medical practitioners in predicting the stage of the disease. 
	
\end{abstract}
		
		\begin{keyword}
			\texttt Avascular tumor, oxygen distribution, Stokes equation, SIMPLE algorithm, FDM, necrotic region
		\end{keyword}
		
	\end{frontmatter}
	
	
	\section{Introduction}
	Tumor cells are very proliferative, and for their proliferation and survival, cells need a continuous supply of oxygen, glucose, and other nutrients \cite{hanahan2000hallmarks}. At the early stage of development of a tumor, these requirements are met from the surrounding tissues. However, as a tumor grows in size, nutrients supply to the tumor central region gradually reduces. As a result, cells die in large quantity due to starvation, and owing to the lack of proper clearance system, dead cells are gathered in the central region, which is known as necrotic core.

In order to explore the dynamics of tumor growth, researchers mostly relied on experimental approaches have been used \cite{ADAM1989,OfraBenny2019tissue,yamamoto2023metastasis}. As, experimental models are costly, time-consuming, and often fail to explain underlying phenomena of growth, so, mathematical models are used to complement the experimental studies. The outcomes of the models help clinicians and biologists to understand the disease progression and growth factors in detail. Over the years, many mathematical models have been proposed to gain insights into tumor growth \cite{collin2021spatial,1972_greenspan,hasan2018classification,mondal2023enhanced,YADAV2021,yadav2023multiscale}
In 1972, Greenspan \cite{1972_greenspan} proposed a mathematical model of tumor growth. The model based on the reaction-diffusion equation, was developed to investigate oxygen transport in an avascular tumor. \citet{casciari1992mathematical} studied the effects of various essential nutrients, like oxygen, glucose, H+ ions, and extracellular pH on tumor growth. They found that the center of the tumor has a lower level of oxygen and glucose concentrations, and it considerably reduces the proliferation rate. 
In 2002, \citet{byrne2003two} developed a two-phase model to elaborate the bio-physical factors of tumor growth. They considered the tumor as a mixture of two phases: tumor cellular phase and ECM (which includes components other than the tumor cells). In the multiphase model, various physical properties, like cellular stress, phase pressure, cell-cell interaction force, cellular viscosity, and ECM phase properties are generally taken into account in avascular tumor growth models \cite{araujo2005mixture,Byrne2002,preziosi2009multiphase,2023two}. \citet{gcrameshan2020} studied the tumor growth in two-dimension using the multiphase model. They developed a finite element based numerical method to cater with the irregular and asymmetric initial tumor geometries.

Although there is a significant development in the mathematical models to explore the tumor growth, the main focus is on the tumor cellular phase development. Several experimental investigations have showed that the necrotic region is also of vital importance \cite{ADAM1989,OfraBenny2019tissue,yamamoto2023metastasis}. \citet{OfraBenny2019tissue} experimentally showed that the contents of dead cells enhances angiogenesis and proliferation of endothelial cells, induces vasculature, and may increase the cells migration. \citet{ADAM1989} studied the effects of necrotic core in tumor growth. They predicted that tumor growth may be hampered in the presence of necrotic core. Therefore, necrotic core is an important component of a tumor growth, and the attention needs to be paid to explore its dynamics.

In this article, we aim to explore the development process of necrotic core of a tumor. The multiphase one-dimensional (1D) continuum-based approach as proposed by \citet{Byrne2002} is adopted to model the growth of a tumor. The tumor is assumed to compose of two phases: tumor cellular phase and ECM as the other phase. The ECM phase contains the extracellular components and majorly the dead cells. So, without considering a separate phase for dead cells in the model, the development of necrotic core is explored using the dynamics of ECM phase. As the supply of oxygen in tissues depends on the diffusional distance from the blood vessels, the tumor region near the blood vessels is rich in oxygen supply compared to the one far from blood vessels \cite{2011oxygen_drop}. As a result, tumor forms asymmetrical shape due to different oxygen supplies at tumor boundaries \cite{1982muellerl}. The main limitation of the existing models is that these were formulated to capture the tumor growth under uniform oxygen supply through its boundaries which helped to establish an explicit relation between tumor cellular and ECM phases velocities  \cite{2003BREWARD,Byrne2002,byrne2003two,2023two}. On the contrary, for the different oxygen concentrations at tumor boundaries such relation does not hold; therefore, the pressure-velocity equations are coupled. We have adopted the \enquote{Semi-Implicit Method for Pressure-Linked Equations} (SIMPLE) algorithm for velocity and pressure coupling in the tumor growth model. A finite difference based numerical method is used using the SIMPLE algorithm in staggered grid. Furthermore, the model is simulated to investigate the growth of necrotic core with various boundary conditions on oxygen supply. Also, the dynamics of tumor cell phase is examined with the different oxygen concentrations at the boundaries. The simulation results are in good agreement with the experimental findings, and some novel insights are observed from the simulation results.

\section{Methods}
	In this section, multiphase mathematical model for tumor growth is  formulated, and numerical scheme to simulate the model is discussed. 
	\subsection{Model formulation}
	Avascular tumor growth is a complex biological process. A tumor consists of various components like tumor cells, dead cells, extracellular matrix, fibroblasts, and collagen fiber. Usually, cell death occurs in two ways: (i) apoptosis that is known as cellular programmed death, and (ii) necrotic death where cell death occurs due to metabolic stress or due to a shortage of nutrients and oxygen supply
	 \cite{fiers1999more,zong2006necrotic}. In this study, the tumor is assumed to be a mixture of two phases: tumor cellular phase and extracellular matrix (ECM). As the intracellular components of dead cells are released into the ECM, so the ECM phase accounts for the necrotic core. It is also assumed that the cellular phase is a viscous fluid phase and ECM as non-viscous one. This is because of the ECM phase where the time-scale for diffusion is much larger than the time-scale for convection compared to that in cellular phase \cite{byrne2003two}. This is due to the presence of solid components, such as structural proteins in the ECM phase, whereas tumor
	 cells have the cytoskeletal network inside it, which hinders cell motion \cite{lemon2006mathematical}. In general, tumor growth depends on the availability of oxygen, glucose, amino acid, and other essential nutrients. However, in this study, oxygen is considered as the only growth factor of the tumor. 
		
	\begin{figure}[h!]
		\begin{center}
			\includegraphics[scale=0.8]{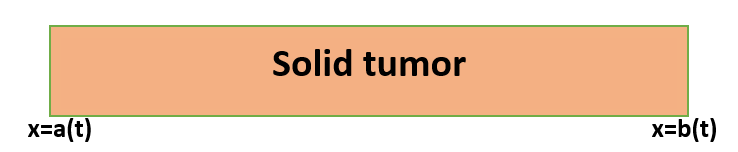}
		\end{center}
		\caption{Schematic diagram of solid tumor.}
		\label{fig:schem_diagram}
	\end{figure}

	\subsection{Governing equations}
	It is assumed that the mixture of two phases is saturated and it constitutes the whole tumor. So, no void condition is imposed as, 
	\begin{equation}
	\alpha+\beta=1,
	\label{eq1}   	
	\end{equation}
	where $\alpha(x,t)$ and $\beta(x,t)$ denote the volume fractions for tumor cellular phase and ECM, respectively. The fluid density $\rho$ is assumed to be constant and equal for both the phases. By applying the conservation of mass as well as of momentum to each phase, the governing equations are obtained as \textcolor{blue}{\cite{Byrne2002}},
	\begin{equation}
	\frac{\partial\alpha}{\partial{t}}+\frac{\partial(\alpha u_{c})}{\partial x}=S_{c},
	\label{eq:mass_conservation1}  	
	\end{equation}
	\begin{equation}
	\frac{\partial\beta}{\partial{t}}+\frac{\partial(\beta u_{w})}{\partial x}=S_{w},
	\label{eq:mass_conservation2}  	  	
	\end{equation}
	\begin{equation}
	\rho\alpha\frac{\partial u_c}{\partial t}=F_{c}+\frac{\partial (\alpha\mathbf{\sigma_{c}})}{\partial x},
	\label{eq:momentum_conservation1}  	  	
	\end{equation}
	\begin{equation}
	\rho\beta\frac{\partial u_w}{\partial t}=F_{w}+\frac{\partial (\beta\mathbf{\sigma_{w}})}{\partial x}.
	\label{eq:momentum_conservation2}  	  	 
	\end{equation}
	Here, $c$ and $w$ in the subscripts denote tumor cellular phase and ECM, respectively. $u$ denotes the velocity, $\sigma$ is the stress tensor, $S$ is the mass source or sink, and $F$ momentum source/sink. $F_c$ represents the force applied by ECM phase on cellular phase. 
	Initially (i.e., at $t=0$), it is assumed that the tumor is not growing, so $u_c=u_w=0$. The initial volume fraction of tumor cellular phase is considered to be $\alpha=\alpha_{0}$.
	
	The oxygen transport is given as \cite{lewin2020three},
	\begin{equation}
	\frac{\partial C}{\partial t}=D\frac{\partial^2 C}{\partial x^2} -\Gamma\alpha C H(C-C_N),
	\label{eq:Oxygen_diffusion}
	\end{equation}
	where $C$ is the oxygen concentration, $D $ is the diffusion coefficient, $C_N$ the necrotic threshold value of oxygen below which tumor cells cannot survive. The tumor cells consume oxygen with a consumption rate $\Gamma$. $H$ is the Heaviside function defined as,
	\begin{align*}
	H(C-C_N)=\begin{cases}
	1, & \text{if $C>C_N$},\\
	0, & \text{otherwise}.
	\end{cases}  
	\end{align*}
	It is considered that the oxygen is available at the tumor boundaries, so the oxygen concentrations are given as,
	\begin{eqnarray}
	C=C_{l}\;\; \mbox{at} \;\; x=a(t),\nonumber \\
	C=C_{r}\;\;\; \mbox{at} \;\; x=b(t).
	\label{eq:oxy_boundary}
	\end{eqnarray}  
	The initial oxygen distribution is considered to be linear over the tumor region $[a, b]=[a_0, b_0]$ at $t=0$, and it is given as,
	\begin{equation}
	C(x)=C_r+(b_0-x)\frac{C_l-C_r}{b_0-a_0} \;\; \text{for}\;\; a_0\leq x\leq b_0.
	\label{eq:oxygen_initialization}
	\end{equation}
	
	\subsubsection{Calculation of mass source/sink terms} 
	The volume fraction ($\alpha$) of tumor cells increases due to cell division and decreases with cell death. So, the source term is defined as \textcolor{blue}{\cite{Byrne2002}},
	\begin{equation}
	S_{c}=\frac{\alpha\beta S_0 C}{1+S_1C}-\frac{S_2+S_3C}{1+S_4C}\alpha,
	\label{eq:mass_source}  	
	\end{equation}
	where the first term on the right-hand-side accounts for the  cell proliferation under nourished conditions with parameters $S_0$ and $S_1$, and the last-term for necrotic cell death due to shortage of oxygen supply with parameters $S_2$, $S_3$, and $S_4$. The deceased cells are thought to dissolve into the ECM and are ready for mitosis. One can note that if the relation $S_2<\frac{S_3}{S_4}$ holds, cell proliferation increases with an increase in $C$ and cell death increases with a decrease in $C$ \cite{Byrne2002}. This restriction is taken into account to investigate the tumor growth dynamics in the present study. However, if this restriction does not hold, tumor may shrink in size.

	As the total mass is conserved, any loss/gain in the volume of one phase via source/sink is balanced by an equal volume change in the other phase. This can be incorporated with the condition
	\begin{equation}
	S_{c}+S_w=0.
	\label{eq:No_void}
	\end{equation}

	\subsubsection{Calculation of momentum source/sink terms} 
	The momentum source/sink terms $F_c$ and $F_w$  are computed using the relations as follows \textcolor{blue}{\cite{2013HUBBARD}},  
	\begin{align}
	F_c=p_c\frac{\partial \alpha}{\partial x}+K \alpha\beta(u_w-u_c),\nonumber \\
	F_w=p_w\frac{\partial \beta}{\partial x}-K \alpha\beta(u_w-u_c),
	\label{eq:momentum_source}
	\end{align}
	where $p_c$ and $p_w$ are pressures in cellular and ECM phases, respectively. $K$ is the drag coefficient associated with the relative movement between the phases. 
	
	Let $\mu_{c}$ be the viscosity of tumor cellular phase. The stresses are given as \textcolor{blue}{\cite{Byrne2002}},
	\begin{align}
	\sigma_{c}&=-p_{c}+2\mu_{c}\frac{\partial u_c}{\partial x},\nonumber\\
	\sigma_{w}&=-p_w, \text{ with }p_c=p_w+\Sigma_{c}.
	\label{eq:stress}
	\end{align}  
	Here, $\Sigma_{c}$ is the cell-cell interaction force. In healthy tissue, cells are sparsely distributed, and hence $\Sigma_{c}$ is negligible. Due to the rapid proliferation, tumor cells come in their close contact, which results in a net cell-cell interaction force. If $0<\alpha<\alpha_{min}$,  the cells are sparsely distributed and do not interact with one another. For $\alpha_{min}<\alpha<\alpha^{*}$, $\Sigma_c <0$, which signifies that the cell experiences attraction force. For $\alpha>\alpha^{*}$, the tumor cell phase exerts a net repulsive force. So, $\Sigma_{c}$ is calculated as \textcolor{blue}{\cite{Byrne2002}},
	\begin{align}
	\Sigma_c=\begin{cases}
	\gamma\frac{(\alpha-\alpha^*)}{(1-\alpha)^2}, & \text{$\alpha>\alpha_{min}$},\\
	0, & \text{otherwise},
	\end{cases}
	\label{eq:cell_cell_interaction}
	\end{align}
	where $\gamma$ is the tension force constant. 
	\subsection{Boundary propagation tracking equations}
	We assume that tumor boundaries propagate with the tumor cellular phase velocity. So, the equations are given as,
	\begin{eqnarray}
	\frac{da}{dt}=u_{c}(a), \\
	\frac{db}{dt}=u_{c}(b).
	\label{eq:boundary_propagation}
	\end{eqnarray} 
	
	\subsection{Model simplification}	
	The momentum equations \eqref{eq:momentum_conservation1}--\eqref{eq:momentum_conservation2} (using Eqs. \eqref{eq:momentum_source} and \eqref{eq:stress}) reduce to
	
	\begin{equation}
	\rho\alpha\frac{\partial u_c}{\partial t}=-\alpha\frac{\partial p_c}{\partial x}+2\mu_{c}\frac{\partial (\alpha\frac{\partial u_c}{\partial x})}{\partial x}+K\alpha\beta(u_w-u_c),
	\label{eq:uc_momentum}
	\end{equation} 
	\begin{equation}
	\rho\beta\frac{\partial u_w}{\partial t}=-\beta\frac{\partial p_w}{\partial x}-K\alpha\beta(u_w-u_c) 
	\label{eq:uw_momentum}
	\end{equation}
	with $p_c=p_w+\Sigma_{c}$. 
	We assume that the stresses are zero at the boundaries and are given as,
	\begin{align}
	\sigma_{c}=\sigma_{w}=0 \;\;\; \mbox{at} \;\;\;\; x=a(t) \;\; \text{and} \;\; b(t), \nonumber \\ 
	2\mu_{c}\frac{\partial u_c}{\partial x}=\Sigma_{c}\;\;\; \mbox{at} \;\;\;\; x=a(t)\;\; \text{and}\;\; b(t).
	\label{eq:stress_zero}
	\end{align}

	\subsection{Non-dimensionalized model}
	Let $R_m$ be the maximal tumor length, $\hat{C}$ be the characteristic oxygen concentration (which is taken as the oxygen concentration at the left boundary of tumor ($C_{l}$)), and
	the characteristic time $\hat{t}=\dfrac{1+S_1\hat{C}}{S_0\hat{C}}$, representing the time scale of proliferation rate under a well-nourished condition \cite{2023two}.
	 So, the dimensionless variables are given as, 
	\begin{equation}
	x^{'}=\frac{x}{R_m}, ~C^{'}=\frac{C}{\hat{C} },\text{ and } t^{'}=\frac{t}{\hat{t}}.
	\label{eq:non_dim_para}
	\end{equation}
	The parameters and variables are made dimensionless in the following way: $a^{'}=\frac{a}{R_m}$, $b^{'}=\frac{b}{R_m}$, $a_0^{'}=\frac{a_0}{R_m}$, $b_0^{'}=\frac{b_0}{R_m}$, $\alpha=\alpha^{'}$, $\beta=\beta^{'}$, $u_c=\frac{R_m u_c^{'}}{\hat{t}}$, $u_w=\frac{R_m u_w^{'}}{\hat{t}}$, $S_1^{'}=S_1 \hat{C}$, $S_2^{'}=S_2 \hat{t}$, $S_4^{'}=S_4 \hat{C}$, $S_3^{'}=S_3 \hat{t} \hat{C}$, $p_c^{'}=\frac{\hat{t}^2p_c}{R_m^2\rho}$, $p_w^{'}=\frac{\hat{t}^2p_w}{R_m^2\rho}$, $\mu_{c}^{'}=\frac{\mu_{c}\hat{t}}{R_m^2\rho}$, $K^{'}=\frac{K\hat{t}}{\rho}$, $\gamma^{'}=\frac{\hat{t}^2\gamma}{R_m^2\rho}$, $D^{'}=\frac{D \hat{t}}{R_m^2}$, $\Gamma^{'}=\hat{t}\Gamma$, $C_N=C_N^{'}\hat{C}$, and $C_r^{'}=\frac{C_r}{\hat{C}}$.

	Then non-dimensional mass conservation equations are obtained as (dropping dashes for convenience),
	\begin{equation}
	\frac{\partial\alpha}{\partial{t}}+\frac{\partial(\alpha u_{c})}{\partial x}=\frac{(1+S_1)\alpha\beta C}{1+S_1C}-\frac{S_2+S_3C}{1+S_4C}\alpha=S_c,
	\label{eq:Non_mass_alpha}	
	\end{equation}
	
	\begin{equation}
	\frac{\partial\beta}{\partial{t}}+\frac{\partial(\beta u_{w})}{\partial x}=-\frac{(1+S_1)\alpha\beta C}{1+S_1C}+\frac{S_2+S_3C}{1+S_4C}\alpha=-S_c.
	\label{eq:Non_mass_beta}  	  	
	\end{equation}
	The non-dimensional momentum equations (using the relation  $p_c=p_w+\Sigma_{c}$) are given as,
	\begin{equation}
	\alpha\frac{\partial u_c}{\partial t}=-\alpha\frac{\partial p_c}{\partial x}+2\mu_{c}\frac{\partial (\alpha\frac{\partial u_c}{\partial x})}{\partial x}+K\alpha\beta(u_w-u_c),
	\label{eq:Non_dim_uc}
	\end{equation}
	
	\begin{equation}
	\frac{\partial u_w}{\partial t}=-\frac{\partial p_c}{\partial x}+\frac{\partial \Sigma_c}{\partial x}-K\alpha(u_w-u_c).
	\label{eq:Non_dim_uw} 
	\end{equation}   
	The zero stress conditions at boundaries in the non-dimensional form are given as,
	\begin{align}
	\sigma_{c}=\sigma_{w}=0 \;\;\; \mbox{at} \;\;\;\; x=a(t) \;\; \text{and} \;\; b(t), \nonumber \\ 
	2\mu_{c}\frac{\partial u_c}{\partial x}=\Sigma_{c}\;\;\; \mbox{at} \;\;\;\; x=a(t)\;\; \text{and}\;\; b(t).
	\label{eq:stress_zero_non_dimesional}
	\end{align}
	Velocities and cellular volume fraction at initial time ($t=0$) are given as,
	\begin{equation}\label{initial_u_and_volume}
	u_c=u_w=0\;\;\text{and}\;\ \alpha=\alpha_{0}.
	\end{equation}
	
	The non-dimensional oxygen diffusion equation becomes
	\begin{equation}
	\frac{\partial C}{\partial t}=D\frac{\partial^2 C}{\partial x^2} -\Gamma\alpha C H(C-C_N)
	\label{eq:Non_dim_oxy}
	\end{equation}
	with boundary conditions
	\begin{eqnarray}
	C=C_l \;\; \mbox{at} \;\; x=a(t),\nonumber \\
	C=C_{r}\;\;\; \mbox{at} \;\; x=b(t),
	\label{eq:non_dim_oxygen_boundary}
	\end{eqnarray}  
	and initial concentration distribution profile
	\begin{equation}
	C(x)=C_r+(b_0-x)\frac{C_l-C_r}{b_0-a_0}\quad\text{ for }\quad a_0\leq x\leq b_0.
	\label{eq:non_dim_oxygen_initialization}
	\end{equation}
	The non-dimensional boundary propagation equations can be written as,
	\begin{eqnarray}
	\frac{da}{dt}=u_{c}(a), \\
	\frac{db}{dt}=u_{c}(b).
	\label{eq:non_dim_boundary_propagation}
	\end{eqnarray}

	\subsection{Numerical method}
\begin{figure}
		\begin{center}
			\begin{tikzpicture}[xscale = 1.5]
			
			\draw [-] (-5,0) -- (5,0);
			\draw (-5,-0.2) -- (-5,0.2);
			\draw (-4.5,-0.2)--(-4.5,0.2);
			\draw (-4,-0.2)--(-4,0.2);
			\draw (-3.5,-0.2)--(-3.5,0.2);

			\draw (1,-0.2) -- (1,0.2);
			\draw (0.5,-0.2) -- (0.5,0.2);
			\draw (0,-0.2) -- (0,0.2); 
			\draw (-0.5,-0.2) -- (-0.5,0.2); 
			\draw (-1,-0.2) -- (-1,0.2); 
			\draw (3.5,-0.2)--(3.5,0.2);
			\draw (4,-0.2)--(4,0.2);
			\draw (4.5,-0.2)--(4.5,0.2);
			\draw (5,-0.2) -- (5,0.2);
			
			\fill (0,0) circle (.07);
			\fill (5,0) circle (.07); 
			\fill (-5,0) circle (.07); 
			\fill (4,0) circle (.07); 
			\fill (-4,0) circle (.07);	
			\fill (1,0) circle (.07); 
			\fill (-1,0) circle (.07);
			\fill (3,0) circle (.07); 
			\fill (-3,0) circle (.07);
			\node [below] at (-5,-0.32) {$x_0 $};
			\node [above] at (-5,0.2) {$\mathbf{V}$};
			
			\node [below] at (-4.5,-0.32) {$x_1$};
			\node [above] at (-4.5,0.2) {$\mathbb{F}$};
			
			\node [below] at (-4,-0.32) {$x_2$};
			\node [above] at (-4,0.2) {$\mathbf{V}$};
			
			\node [below] at (-3.5,-0.32) {$x_3$};
			\node [above] at (-3.5,0.2) {$\mathbb{F}$};
			
			\node [below] at (-1,-0.32) {$x_{i-2} $};
			\node [above] at (-1,0.2) {$\mathbf{V}$};
			
			\node [below] at (-0.5,-0.32) {$x_{i-1} $};
			\node [above] at (-0.5,0.2) {$\mathbb{F}$};
			
			\node [below] at (0,-0.32) {$x_i $};
			\node [above] at (0,0.2) {$\mathbf{V}$};

			\node [below] at (0.5,-0.32) {$x_{i+1} $};
			\node [above] at (0.5,0.2) {$\mathbb{F}$};
			
			\node [below] at (1,-0.32) {$x_{i+2} $};
			\node [above] at (1,0.2)  {$\mathbf{V}$};

			\node [below] at (3.5,-0.6) {$x_{N-4}$};
			\node [above] at (3.5,0.2) {$\mathbb{F}$};
			
			\node [below] at (4,-0.25) {$x_{N-3}$};
			\node [above] at (4,0.2) {$\mathbf{V}$};

			\node [below] at (4.5,-0.5) {$x_{N-2}$};
			\node [above] at (4.5,0.2) {$\mathbb{F}$};
			
			\node [below] at (5,-0.32) {$x_{N-1}$};
			\node [above] at (5,0.2) {$\mathbf{V}$};
			
			\end{tikzpicture}
		\end{center}
		\caption{Schematic diagram of the grid points and the staggered arrangement of variable on the computational domain. Here, $\mathbb{F}$ denotes the scaler quantities $\{C,p_c,\alpha\}$  and $\mathbf{V}$ denotes the velocities $\{u_c,u_w\}$.}
		\label{fig:fig11}
	\end{figure}
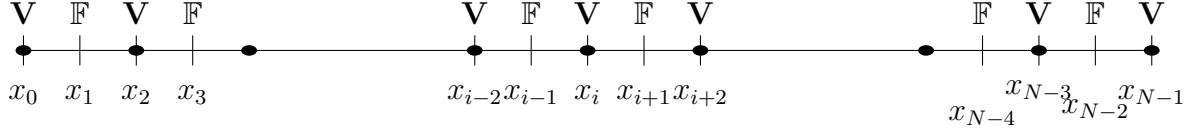

	This section presents the numerical procedure to solve the mathematical model [Eqs. \eqref{eq:Non_mass_alpha} -- \eqref{eq:Non_dim_oxy}]. The governing equations are discretized in a staggered framework. The staggered grid is chosen in such a way that the velocity $u_c$ is on the boundary nodes as depicted in Fig. \ref{fig:fig11}. The domain $[a,b]$ is discretized with step size $\Delta x=\left(\frac{b-a}{N-1}\right)$, where $N$ is the number of grid points. The spatial grid points are numbered as $i=0,1,\ldots,N-1$
	
	At the interior nodes, a central difference scheme is used for spatial discretization. While for boundary nodes, a special treatment is adopted as discussed below. The first-order forward difference scheme is adopted for temporal derivatives.

	\subsection{Discretization of momentum conservation equations}
	The discretized form of the Eq. \eqref{eq:Non_dim_uw} is given as, 
	\begin{equation}
	u_{w,i}^{n+1}=\frac{1}{1+A_1 \alpha_i}(u_{w,i}^n+A_1 \alpha_iu_{c,i}^{n+1}-B_1(p_{c,i+1}-p_{c,i-1})+B_1(\Sigma_{c,i+1}-\Sigma_{c,i-1})),
	\label{eq:discrete_uw}
	\end{equation}
	where $A_1=K\Delta t,B_1=\frac{\Delta t}{2\Delta x} $, and $\alpha_i=\frac{\alpha_{i+1}+\alpha_{i-1}}{2}$ for $i=2,4,\ldots,N-3$. $u_{c,i}^{n+1}$ and $u_{w,i}^{n+1}$ are the discrete counterparts of $u_c$ and $u_w$ respectively at time $t_{n+1}=(n+1)\Delta t$ for time-step $\Delta t$ and at position $x_i=i\Delta x$.\\
	Using the expression of $u_{w,i}^{n+1}$, the discretized form of Eq. \eqref{eq:Non_dim_uc} can be written as, 
	\begin{multline}
	-\frac{B_1\mu_{c}\alpha_{i-1}}{\Delta x^2\alpha_i}u_{c,i-2}^{n+1}+\left(\frac{A_2}{1+A_1\alpha_i}+\frac{B_1\mu_{c}(\alpha_{i-1}+\alpha_{i+1})}{\Delta x\alpha_i}\right)u_{c,i}^{n+1}-\frac{B_1\mu_{c}\alpha_{i+1}}{\Delta x\alpha_i}u_{c,i+2}^{n+1}\\=-\frac{A_2B_1}{1+A_1\alpha_i}(p_{c,i+1}-p_{c,i-1})+\frac{A_1(1-\alpha_i) }{1+A_1\alpha_i }\left(u_{w,i}^n+B_1(\Sigma_{c,i+1}-\Sigma_{c,i-1})\right),
	\label{eq:discrete_uc}
	\end{multline}
	where $A_2=1+A_1$.\\
	At the left boundary (i.e., for $i=0$), the discretized form of Eq. \eqref{eq:stress_zero_non_dimesional} is obtained as,
	\begin{equation}
	u_{c,0}=u_{c,2}-\frac{\Delta x \Sigma_{c,0}}{\mu_{c}}.
	\label{eq:uc_at0}
	\end{equation}
	At the right boundary (i.e., for $i=N-1$), the discretized form of Eq. \eqref{eq:stress_zero_non_dimesional} is obtained as,
	\begin{equation}
	u_{c,N-1}=u_{c,N-3}+\frac{\Delta x \Sigma_{c,N-1}}{\mu_{c}}.
	\label{eq:uc_atN-1}
	\end{equation}
	The boundary information required in the Eq. \eqref{eq:discrete_uc} are supplied by Eqs. \eqref{eq:uc_at0} and \eqref{eq:uc_atN-1}.

	\subsection{Discretization of mass conservation equation}
	Mass conservation Eq. \eqref{eq:Non_mass_alpha} is discretized using FTCS scheme. The discretized equations are given as,
	\begin{equation}
	\alpha_i^{n+1}=\alpha_i^n +\Delta t S_{c,i}^n -(f_{i+1}^n-f_{i-1}^n)B_1,
	\label{eq:discrete_mass}
	\end{equation}
	where $f_{i+1}^n=u_{c,i+1}^n \frac{(\alpha_{i+2}^n+\alpha_{i}^n)}{2}$ and $f_{i-1}^n=u_{c,i-1}^n \frac{(\alpha_{i}^n+\alpha_{i-2}^n)}{2}$ with $f_0^n=u_{c,0}^n\alpha_0^n$ and $f_{N-1}^n=u_{c,N-1}^n\alpha_{N-1}^n$ for $i=1,3,\ldots,N-2$.\\
	Since, the explicit scheme is chosen for mass conservation equation, by following the Fourier stability analysis \cite{cfdbook},  $\Delta t$ is chosen as,
	\begin{align*}
		\Delta t \le \max_i\left(\dfrac{4\Delta x|u_{c,i+1}-u_{c,i-1}|}{u_{c,i+1}^2+u_{c,i-1}^2}\right)\;\text{for}\; i=1,3,\dots, N-2.
		\end{align*}
	Once $\alpha$ is obtained, $\beta$ can be calculated from $\alpha+\beta=1$.
	
	\subsection{Discretization of oxygen diffusion equation} 
	The oxygen diffusion equation \eqref{eq:Non_dim_oxy} is discretized as,	\begin{equation}\label{eq:discrete_oxygen}
	\frac{D_1}{4}C_{i-2}^{n+1}-\left(1+\frac{D_1}{2}+\Gamma\alpha_i^{n+1}H(C_i^n-C_N)\right)C_i^{n+1}+ \frac{D_1}{4}C_{i+2}^{n+1}=-C_i^n,
	\end{equation}
	where $i=3,5,\ldots,(N-4)$ \text{and} $D_1=\frac{D \Delta t}{\Delta x^2}$. \\
	Note that the Eq. \eqref{eq:discrete_oxygen} is obtained by using central differencing with spatial step size $2\Delta x$. For $i=1$, the central differencing is used with spatial size $\Delta x$. The resulting equation involves $C_2$, which is calculated by taking the average of $C_1$ and $C_3$. The equation for $i=1$ is obtained as,
	
	\begin{equation}\label{eq32}
	-\left(1+\frac{3D_1}{2}+\Gamma \alpha_1^{n+1}H(C_1^n-C_N)\right)C_1^{n+1}+ \frac{D_1}{2}C_{3}^{n+1}=-C_1^n-D_1 C_0^{n+1}.
	\end{equation}
	By following the similar approach, the equation for $i=N-2$ is obtained as,  
	\begin{equation}\label{eq33}
	\frac{D_1}{2}C_{N-4}^{n+1} -\left(1+\frac{3D_1}{2}+\Gamma\alpha_{N-2}^{n+1}H(C_{N-2}^n-C_N)\right)C_{N-2}^{n+1}=-C_{N-2}^n-D_1 C_{N-1}^{n+1}.
	\end{equation}

	\subsection{SIMPLE algorithm for momentum equations}
	
	In Eq. \eqref{eq:discrete_uc}, the pressure and velocity are coupled, so SIMPLE algorithm \cite{patankar} is adopted to solve for velocity. 
	The algorithm is as follows.
	
	Let us write the Eq. \eqref{eq:discrete_uc} in the form
	\begin{equation}\label{eq:new_uc_discrete}
	a_i u_{c,i}^{n+1}=\sum a_{nb} u_{c,nb}^{n+1} -\frac{A_2 B_1}{1+A_1\alpha_i}(p_{c,i+1}-p_{c,i-1})+\phi(\alpha_i,\Sigma_{c,i}),
	\end{equation}
	where $u_{c,nb}$ are the neighborhood velocities  around the grid point $i$.
	
	Let $p_c^*$ be the guessed pressure field and take $p_c=p_c^*$ in Eqs. \eqref{eq:new_uc_discrete}. The resulted system of equations is obtained as,
	\begin{equation}\label{eq:pc_uc_statisfied}
	a_i u_{c,i}^{*n+1}=\sum a_{nb} u_{c,nb}^{*n+1} -\frac{A_2 B_1}{1+A_1\alpha_i}(p_{c,i+1}^*-p_{c,i-1}^*)+\phi(\alpha_i,\Sigma_{c,i}).
	\end{equation}
	Here, $u_{c,i}^*$ can be called the pseudo velocity field based on a guessed pressure field $p_c^{*}$.

	If $u_c^{'}$ and $p_c'$ are the velocity and pressure correction terms, then 
	\begin{align}
	u_c&=u_c^*+u_c^{'},	\label{eq:velo_corrected}\\
	p_c&=p_c^*+p_c^{'}.
	\label{eq:pressure_corrected}
	\end{align}
	Subtracting Eq. \eqref{eq:pc_uc_statisfied} from \eqref{eq:new_uc_discrete}, we have
	
	\begin{equation}\label{eq37}
	a_i u_{c,i}^{' n+1}=\sum a_{nb} u_{c,nb}^{' n+1} -\frac{A_2 B_1}{1+A_1\alpha_i}(p_{c,i+1}^{'}-p_{c,i-1}^{'}).
	\end{equation}
    The omission of the term $\sum a_{nb} u_{c,nb}^{' n+1}$ in deriving the pressure correction equations is of no consequence as far as the final converged results are concerned. Detailed information on this is available in the book of S.V. Patankar \cite{patankar}. So by dropping the term $\sum a_{nb} u_{c,nb}^{' n+1}$, Eq. \eqref{eq37} can be written as,
	
	\begin{equation}\label{eq:corrected_velocity_uc}
	u_{c,i}^{' n+1}=-\frac{A_2 B_1}{a_i(1+A_1\alpha_i)}(p_{c,i+1}^{'}-p_{c,i-1}^{'}),
	\end{equation}
	where $a_i=\left(\frac{A_2}{1+A_1\alpha_i}+\frac{B_1\mu_{c}(\alpha_{i-1}+\alpha_{i+1})}{\Delta x\alpha_i}\right)$, $\alpha_i=\frac{\alpha_{i+1}+\alpha_{i-1}}{2} \;\; \text{for}\;\;\;  i=2,4,\ldots,N-3$.  $nb$ denotes neighboring points of $i$.
	Eq. \eqref{eq:corrected_velocity_uc} is called the velocity-correction formula. Therefore, Eq. \eqref{eq:velo_corrected} can be rewritten as,
	\begin{equation}\label{eq:correct_velocity_uc}
	u_{c,i}^{n+1}=u_c^*-\frac{A_2 B_1}{a_i(1+A_1\alpha_i)}(p_{c,i+1}^{'}-p_{c,i-1}^{'}).
	\end{equation}

	Upon summing up the mass conservation Eqs. \eqref{eq:Non_mass_alpha} and \eqref{eq:Non_mass_beta} and using $\alpha+\beta=1$, the pressure correction equation can be obtained from
	\begin{equation}\label{eq:continuity}
	\frac{\partial (\alpha u_c +\beta u_w)}{\partial x}=0.
	\end{equation}
	By using the central difference scheme in Eq. \eqref{eq:continuity} and by substituting the values of $u_c$ from \eqref{eq:discrete_uc} and $u_w$ from \eqref{eq:discrete_uw}, the pressure correction equation is obtained as, 
	\begin{equation}\label{eq:pressure_correction}
	B_1(H_{i-1}-E_{i-1})p_{c,i-1}^{'}-B_1(H_{i-1}-E_{i-1}+H_{i+1}-E_{i+1})p_{c,i+1}^{'}+B_1(H_{i+1}-E_{i+1})p_{c,i+3}^{'}=Q_{i-1}-Q_{i+1},
	\end{equation}
	where 
	$H_{i-1}=-\frac{A_2^2 \alpha_{i-1}}{a_{i-1}(1+A_1 \alpha_{i-1})}, E_{i-1}=\frac{\beta_{i-1}}{1+A_1\alpha_{i-1}}$, and\\
	$Q_{i-1}=\frac{\beta_{i-1}}{1+A_1 \alpha_{i-1}}\left(u_{w,i-1}^n+B_1(\Sigma_{c,i}-\Sigma_{c,i-2})-B_1(p_{c,i}^*-p_{c,i-2}^*)\right)+\frac{\alpha_{i-1}A_2}{1+A_1 \alpha_{i-1}}u_{c,i-1}^{* n+1}$ for $i=3,5,\ldots,N-4.$

	For $i=1$ and $i=N-2$, the pressure correction equations are obtained as,
	
	\begin{align}\label{eq:pressure_correction_at1}
	\left(\frac{2\beta_0}{1+A_1\alpha_{0}}+E_2-H_2-\frac{\alpha_{0}A_2^2}{a_2(1+A_1\alpha_{0})(1+A_1\alpha_{2})}\right)B_1 p_{c,1}^{'n+1}\nonumber\\ 
	+ \left(H_2-E_2+\frac{\alpha_{0}A_2^2}{a_2(1+A_1\alpha_{0})(1+A_1\alpha_{2})}\right)B_1 p_{c,3}^{'n+1}\nonumber\nonumber\\ = 
	\frac{2\beta_0}{1+A_1\alpha_{0}}(u_{w,0}^{n}+2B_1\Sigma_{c,1}) - E_2\left(u_{w,2}^n+B_1(\Sigma_{c,3}-\Sigma_{c,1})\right)\nonumber\\ - \left(\frac{\alpha_{2}A_2}{1+A_1\alpha_{2}}-\frac{\alpha_{0}A_2}{1+A_1\alpha_{0}}\right)u_{c,2}^{*}\nonumber\\-\frac{\alpha_{0}A_2}{1+A_1\alpha_{0}}\frac{\Delta x \Sigma_{c,0}}{\mu_{c}}+E_2 B_1 (p_{c,3}^{*}-p_{c,1}^{*})-\frac{\beta_0 B_1}{1+A_1 \alpha_{0}}p_{c,1}^{*},
	\end{align}
	and
	\begin{multline}\label{eq:pressure_correction_atN-2}
	\left(H_{N-3}-E_{N-3}+\frac{\alpha_{N-1}A_2^2}{a_{N-3}(1+A_1\alpha_{N-1})(1+A_1\alpha_{N-3})}\right)B_1p_{c,N-4}^{'n+1}\\
	+ \left(\frac{2\beta_{N-1}}{1+A_1\alpha_{N-1}}+E_{N-3}-H_{N-3}-\frac{\alpha_{N-1}A_2^2}{a_{N-3}(1+A_1\alpha_{N-1})(1+A_1\alpha_{N-3})}\right)B_1p_{c,N-2}^{'n+1}
	=\\-\frac{2\beta_{N-1}}{1+A_1\alpha_{N-1}}(u_{w,N-1}^{n}
	-2B_1\Sigma_{c,N-2})+E_{N-3}\left(u_{w,N-3}^n+B_1(\Sigma_{c,N-2}-\Sigma_{c,N-4})\right)\\
	- \left(\frac{\alpha_{N-1}A_2}{1+A_1\alpha_{N-1}}-\frac{\alpha_{N-3}A_2}{1+A_1\alpha_{N-3}}\right)u_{c,N-3}^{*}\\
	-\frac{\alpha_{N-1}A_2}{1+A_1\alpha_{N-1}}\frac{\Delta x \Sigma_{c,N-1}}{\mu_{c}}-
	E_{N-3} B_1 (p_{c,N-2}^{*}-p_{c,N-4}^{*})
	-\frac{\beta_{N-1} B_1}{1+A_1 \alpha_{N-1}}p_{c,N-2}^{*}.
	\end{multline}
	The system of equations for pressure correction is solved using the tridiagonal matrix algorithm (TDMA). 
	Once the pressure correction is obtained, the pressure is  updated using Eq  \eqref{eq:pressure_corrected}. With this corrected pressure, the velocity $u_{c,i}^{**n+1}$ is obtained from Eq. \eqref{eq:discrete_uc}. Also, it can be obtain $u_{w,i}^{**n+1}$ from Eq. \eqref{eq:discrete_uw} by substituting $u_{c,i}^{n+1}=u_{c,i}^{**n+1}$ and the corrected pressure for $p_{c,i}$. If $u_{c,i}^{**n+1}$ and $u_{w,i}^{**n+1}$ satisfy the discretized version of the continuity equation \eqref{eq:continuity}, the convergence criteria is given by,
	\begin{equation}\label{convergence}
	\alpha_{i+1}u_{c,i+1}^{**n+1}-\alpha_{i-1}u_{c,i-1}^{**n+1}+\beta_{i+1}u_{w,i+1}^{**n+1}-\beta_{i-1}u_{w,i-1}^{**n+1}=0\;\;\text{for}\;\; i=1,3,\ldots,N-2,
	\end{equation}
where $\alpha_{i+1}=\frac{\alpha_{i+2}+\alpha_{i}}{2}$ and $\beta_i=1-\alpha_i$.
The absolute value of the right-hand side of Eq. \eqref{convergence} is achieved during simulation with a tolerance value $tol$ (in this study $tol=10^{-10}$).

	The SIMPLE algorithm is followed in the following sequence:
	\begin{enumerate}
		\item Guess a pressure field $p_c^{*}$.
		\item Solve the momentum Eq. \eqref{eq:new_uc_discrete} to obtain $u_c^{*}$.
		\item Solve for pressure correction $p_c^{'}$ using Eq. \eqref{eq:pressure_correction}.
		\item  Correct the pressure field by Eq. \eqref{eq:pressure_corrected}.
		\item Correct the velocity field by Eq. \eqref{eq:velo_corrected}.
		\item Test the convergence using Eq. \eqref{convergence}.
		\item  If solution converges, stop. Else return to step 1 with $p_c$ as the new guessed pressure $p_c^{*}$ and repeat the process until converges.
	\end{enumerate}
		
	\subsection{Algorithm for solving tumor growth model}\label{section:my}
	The governing equations of the present model are solved with the following algorithm.
	
	\begin{itemize}
		\item[Step 1:] Initialization, i.e., at $t=0$,\\
		$\alpha_0=0.8, a_{old}=-R_{old},b_{old}=R_{old},R_{old}=0.05,u_c=0,\text{ and }u_w=0$, and solve the oxygen diffusion equation to obtain the initial concentration distribution.\\
		
		\textbf{For each time-step until $t=T$
			\;\;\; do}
		\item[Step 2:] Solve the momentum equations to get the velocity field using the initial volume fractions.
		\item[Step 3:] Solve mass conservation equations using the velocities obtained in Step 2.
		\item[Step 4:] Update boundaries $a_{new},b_{new}$ by\\
		$a_{new}=a_{old}+u_{c,0}\Delta t$,\\
		$b_{new}=b_{old}+u_{c,N-1}\Delta t$.
		\item[Step 5]: Update the oxygen field.\\
		
		\textbf{end}
		
	\end{itemize}
	
	\section{Results}
	
	\subsection{Experimental validation}
\begin{figure}[h!]
		\centering
		\includegraphics[scale=0.1]{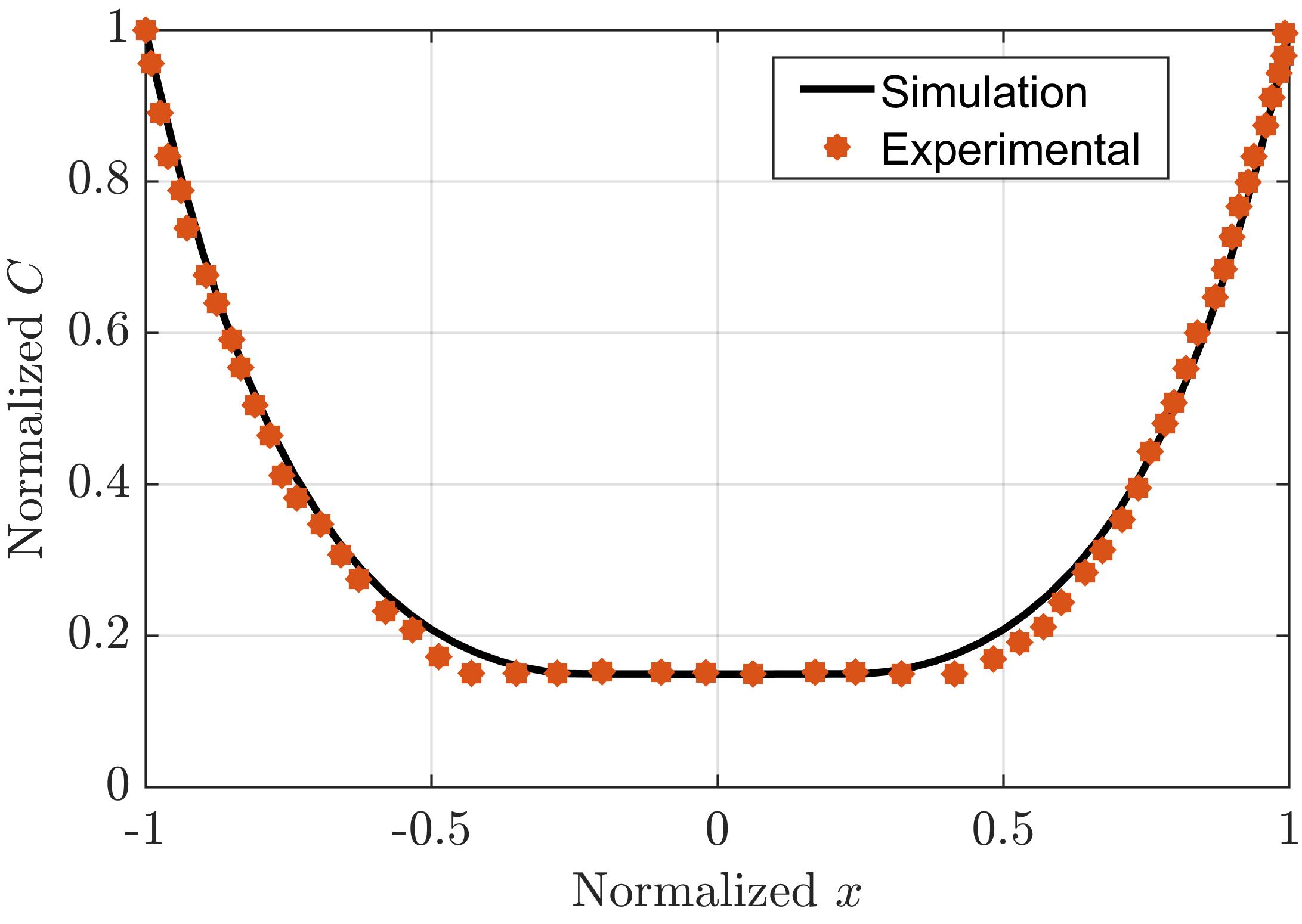}
		\caption{Comparison of predicted results with
			the results of \citet{1982muellerl}.}
		\label{fig:validation}
	\end{figure}

	The numerical results obtained from the present model are validated with the experimental results of \citet{1982muellerl}. They studied the oxygen distribution for various sizes of tumor using oxygen-sensitive microelectrodes. In order to perform numerical simulations, the values of the model parameters are taken from the existing literature and are presented in Table \ref{tab:parameter}. We consider $\alpha=\alpha_{min}=\alpha^{*}=0.8$ as the initial volume fraction \cite{2023two}. So, no cell-cell interaction is considered at the beginning of tumor growth (follows from Eq. \eqref{eq:cell_cell_interaction}). Let $R_0=b_0-a_0$ be the initial dimensionless diameter of tumor (i.e., at $t=0$). Experimental study by \citet{diameter} suggests that the maximum diameter of solid tumor is approximately $200 \mu m$ before the angiogenesis process. In this study, the initial tumor diameter is considered to be $20\mu m$ for simulation. Therefore, $R_0=\frac{1}{10} (=\frac{20}{200})$ is chosen, and the model is simulated up to $t=100$. The comparisons are shown in Fig. \ref{fig:validation}. It can be seen that the numerical results are in very good agreement with the experimental data. This comparison ensure that the model is a realistic and reliable one.

	\begin{table}[h!]
		\centering
		\caption{Parameters values.}
		\begin{tabular}{ p{3cm}  p{3cm} p{3cm}  }
			\hline
			
			Parameter & Non-dimensional value & Reference\\
			\hline
			$\Gamma$   &  $0.1$& \cite{2023two} \\
			$D$    &$\dfrac{1}{1600}$ & \cite{2023two} \\
			$C_N$  &$0.1$ & \cite{lewin2020three}\\
			$\mu_c$&1.0 & \cite{Byrne2002}\\
			$\gamma$    &500    &  \cite{2023two} \\
			$K$   & 1600   & \cite{2023two}    \\
			$S_1$& 10   &  \cite{Byrne2002}   \\
			$S_2$&  0.5 & \cite{Byrne2002}\\
			$S_3$& 0.5  & \cite{Byrne2002}\\
			$S_4$& 10  & \cite{Byrne2002}\\
			
			\hline
			
		\end{tabular}

		\label{tab:parameter}
	\end{table}

   \subsection{Effects of oxygen on necrotic core of tumor}
\begin{figure}[h!]\centering
   	\begin{subfigure}{0.5\textwidth}
   		\centering
   		\includegraphics[scale=0.08]{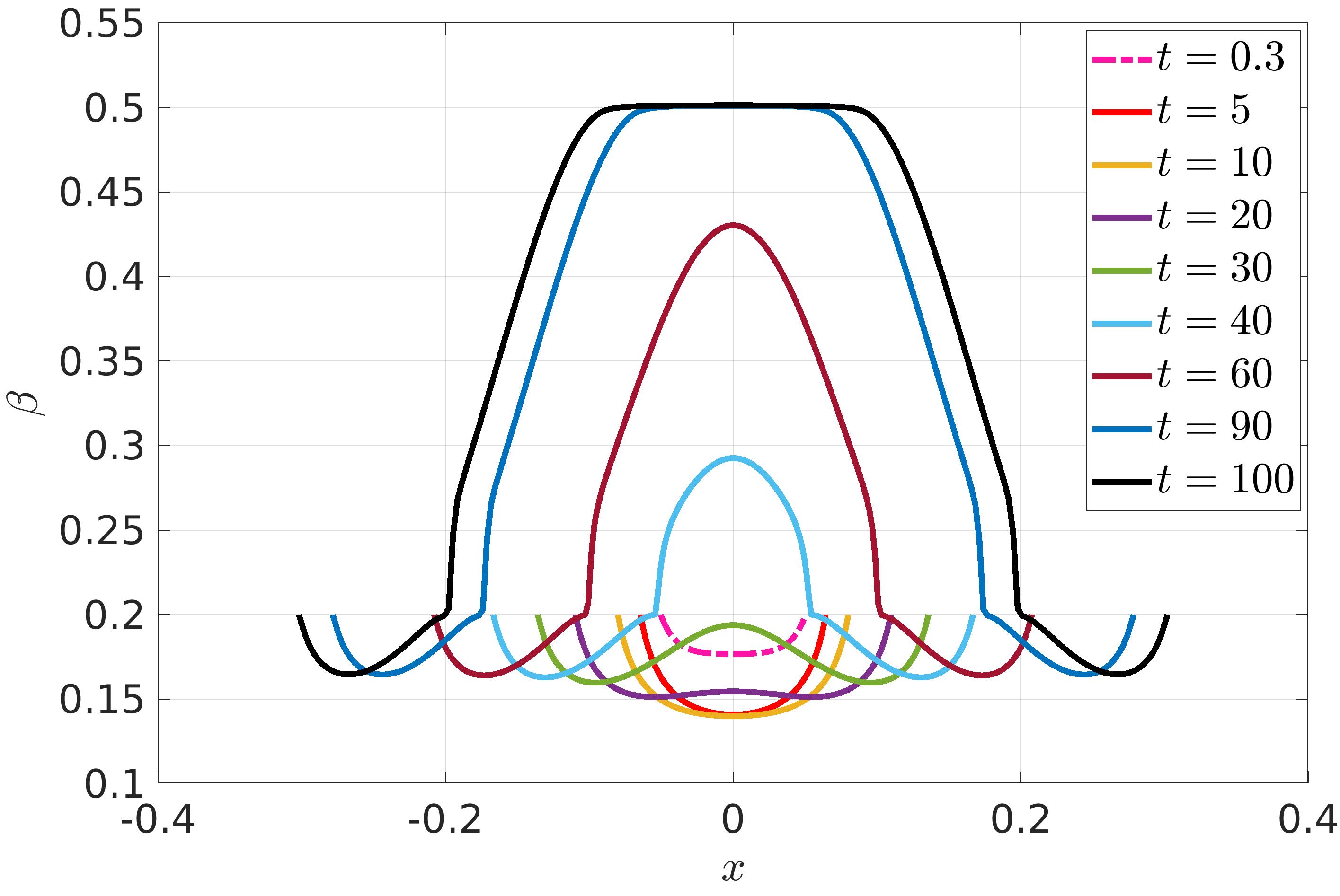}
   		\caption{}
   		\label{fig:ecm}
   	\end{subfigure}	
   	\begin{subfigure}{0.49\textwidth}
   		\centering
   		\includegraphics[scale=0.08]{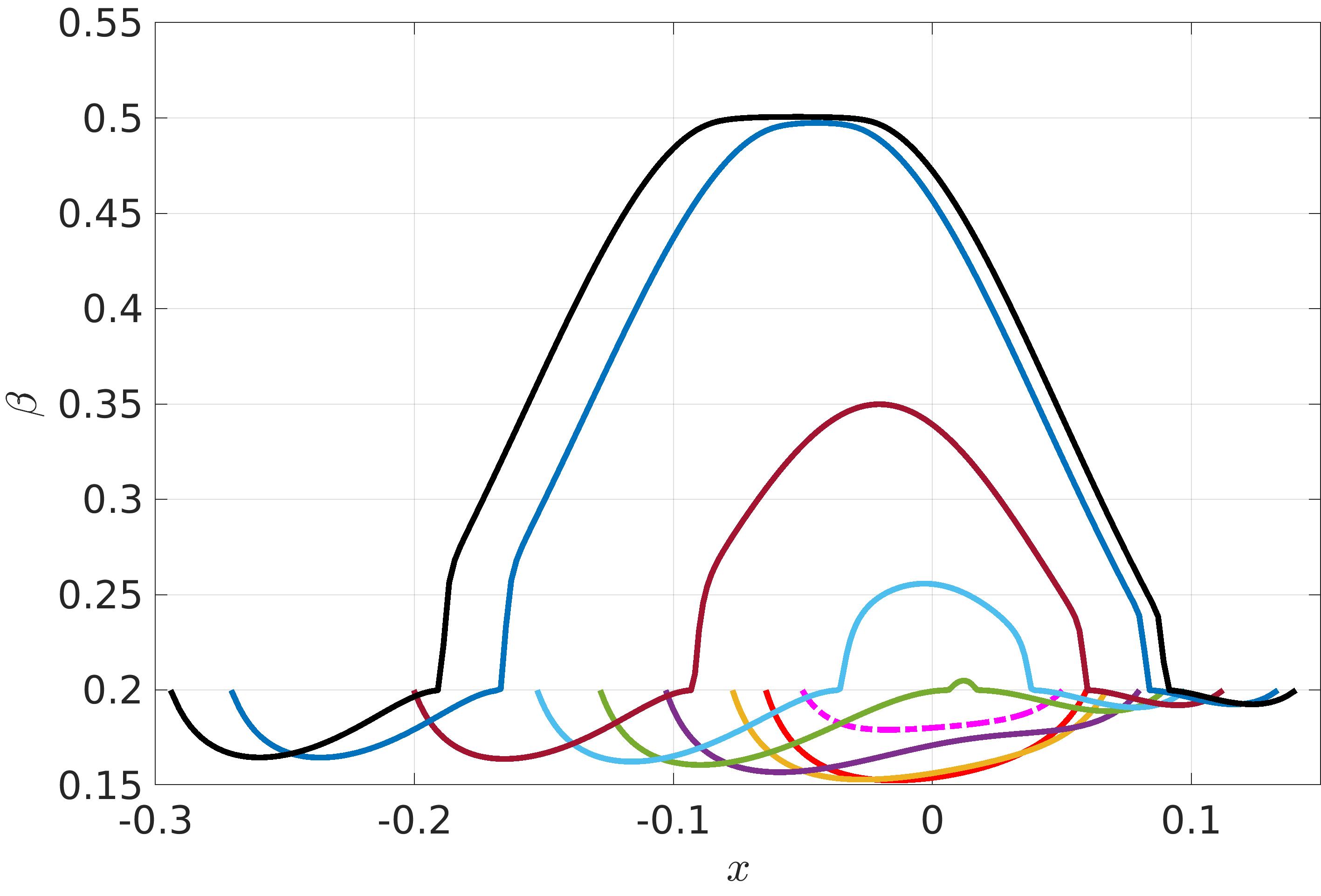}
   		\caption{}		
   		\label{fig:ecm05}		
   	\end{subfigure}		
   	\begin{subfigure}{0.5\textwidth}
   		\centering
   		\includegraphics[scale=0.08]{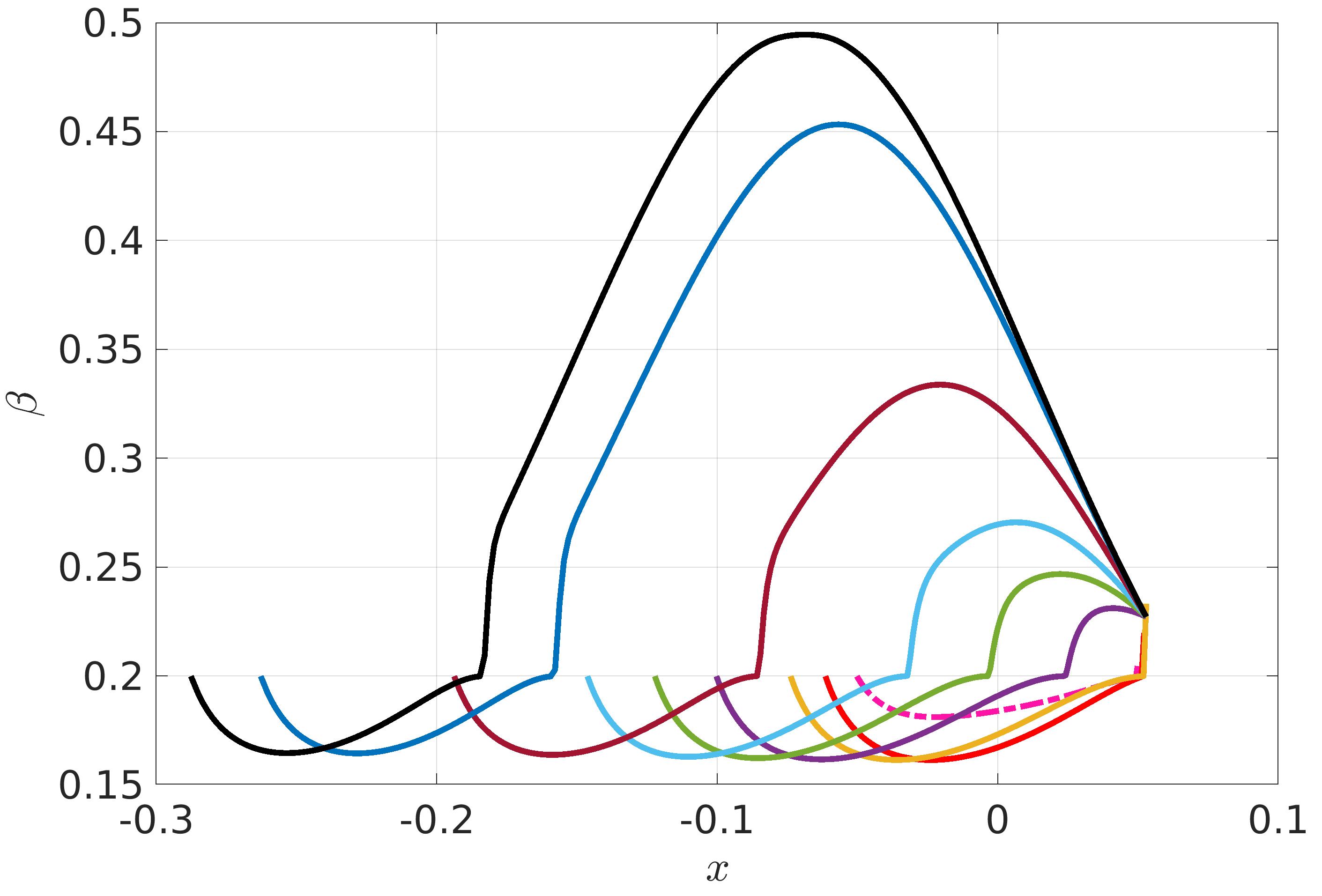}
   		\caption{}		
   		\label{fig:ecm025}
   	\end{subfigure}
   	\begin{subfigure}{0.49\textwidth}
   		\centering
   		\includegraphics[scale=0.08]{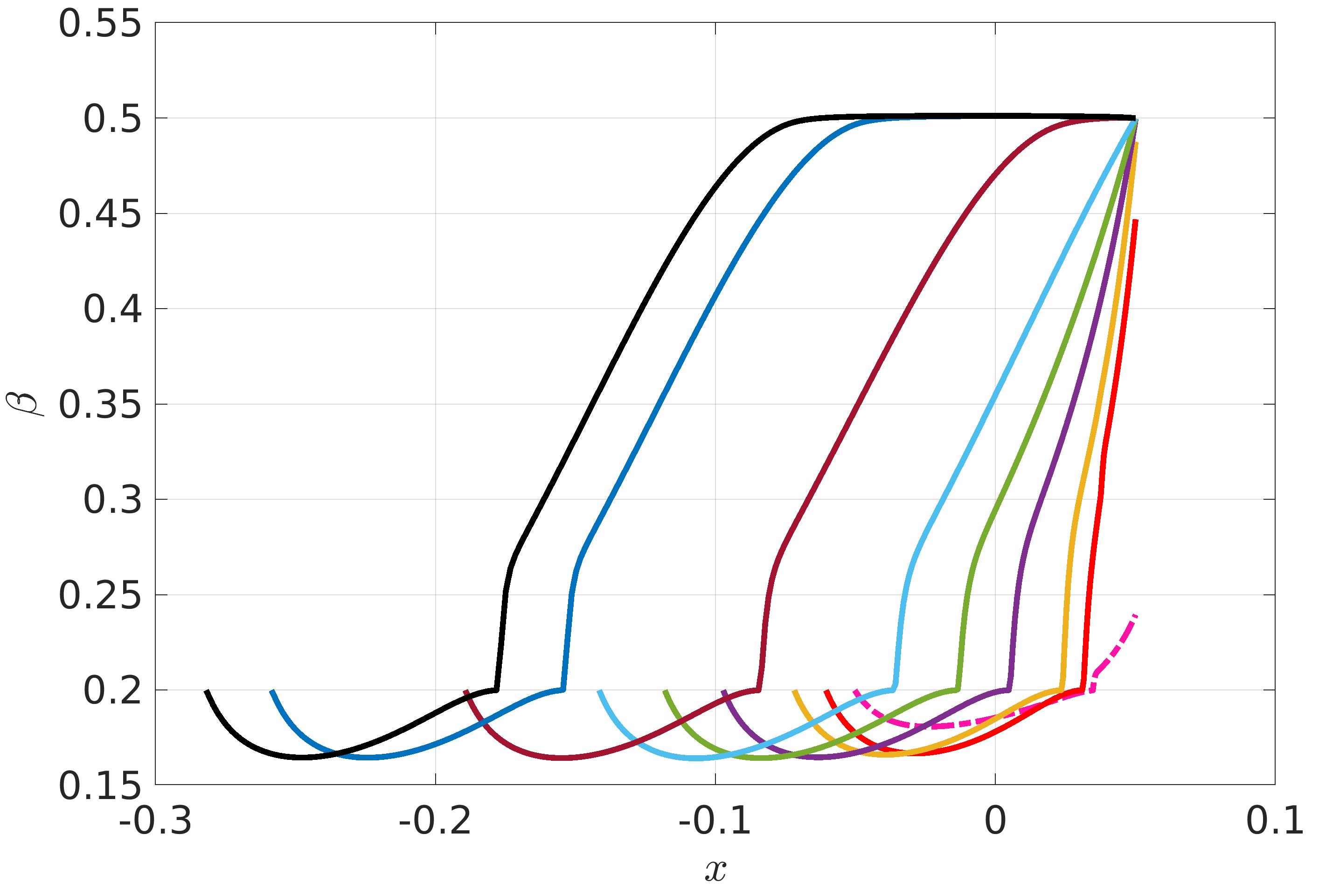}
   		\caption{}		
   		\label{fig:ecm0101}
   	\end{subfigure}
   	\caption{ECM volume fraction in tumor region with oxygen boundary concentrations (a) $C_l=C_r=1$, (b) $C_l=1$,  $C_r=0.5$, (c) $C_l=1$, $C_r=0.25$, and (d) $C_l=1$, $C_r=0.1$.}
   	\label{fig:necrotic}
   \end{figure}

   The growth dynamics of the necrotic region is explored with four different oxygen concentrations at tumor boundaries such as $(a)\; C_l=C_r=1$, $(b)\; C_l=1, C_r=0.5$, $(c)\; C_l=1, C_r=0.25$, and $(d)\; C_l=1, C_r=0.1$, and the results are shown in Fig. \ref{fig:necrotic}. As the dead cells majorly contribute to the ECM phase, without considering the dead phase separately, necrotic core evolution is tracked with ECM volume fraction.
   
   The evolution of necrotic core with equal oxygen concentrations (i.e., $C_l=C_r=1$) at the boundaries is depicted in Fig. \ref{fig:ecm}. The ECM volume fraction decreases initially (i.e., for $t\le30$). It happens due to the proliferation of tumor cells as sufficient oxygen is available in the small-sized tumor. As a result, tumor cellular fraction occupies the ECM space. As the tumor grows in size, tumor cells start to die in the central region. Thereafter, dead cells largely contribute to the increment of ECM volume fraction. The ECM volume faction (or size of the necrotic core) starts to increase in the central region when time $30<t<40$. Owing to the equal oxygen concentrations at boundaries, necrotic core is symmetrical about the tumor center (i.e., $x=0$). 
When $C_l=1$ and $C_r=0.5$, the necrotic core is noticed for $t\ge30$. The key observation in this case is that the necrotic region is no longer symmetrical (Fig. \ref{fig:ecm05}). As time progresses, the necrotic core increases in size and is developed close to the boundary having lower level of oxygen concentration i.e., $C_r=0.5$.
    
As the right boundary oxygen concentration reduces further (i.e., to $C_r=0.25\;\text{and} \;C_r=0.1$), since the beginning, a necrotic core evolves and nears the boundary of reduced oxygen supply (Figs. \ref{fig:ecm025}, \ref{fig:ecm0101}). As the right boundary oxygen concentration decreases, this boundary is occupied with the dead cells. This happens because the tumor cells do not survive due to the oxygen supply shortage. 
    
It can be concluded that the necrotic core is developed in the region with lower oxygen concentration and the morphology of necrotic core changes with the variation of oxygen distribution. Also, necrotic core develops faster with lower level of oxygen concentration.

\subsection{Effects of oxygen supply on tumor cellular volume fraction}
\begin{figure}[h!]
    	\begin{subfigure}{0.49\textwidth}\centering
    		\includegraphics[scale=0.08]{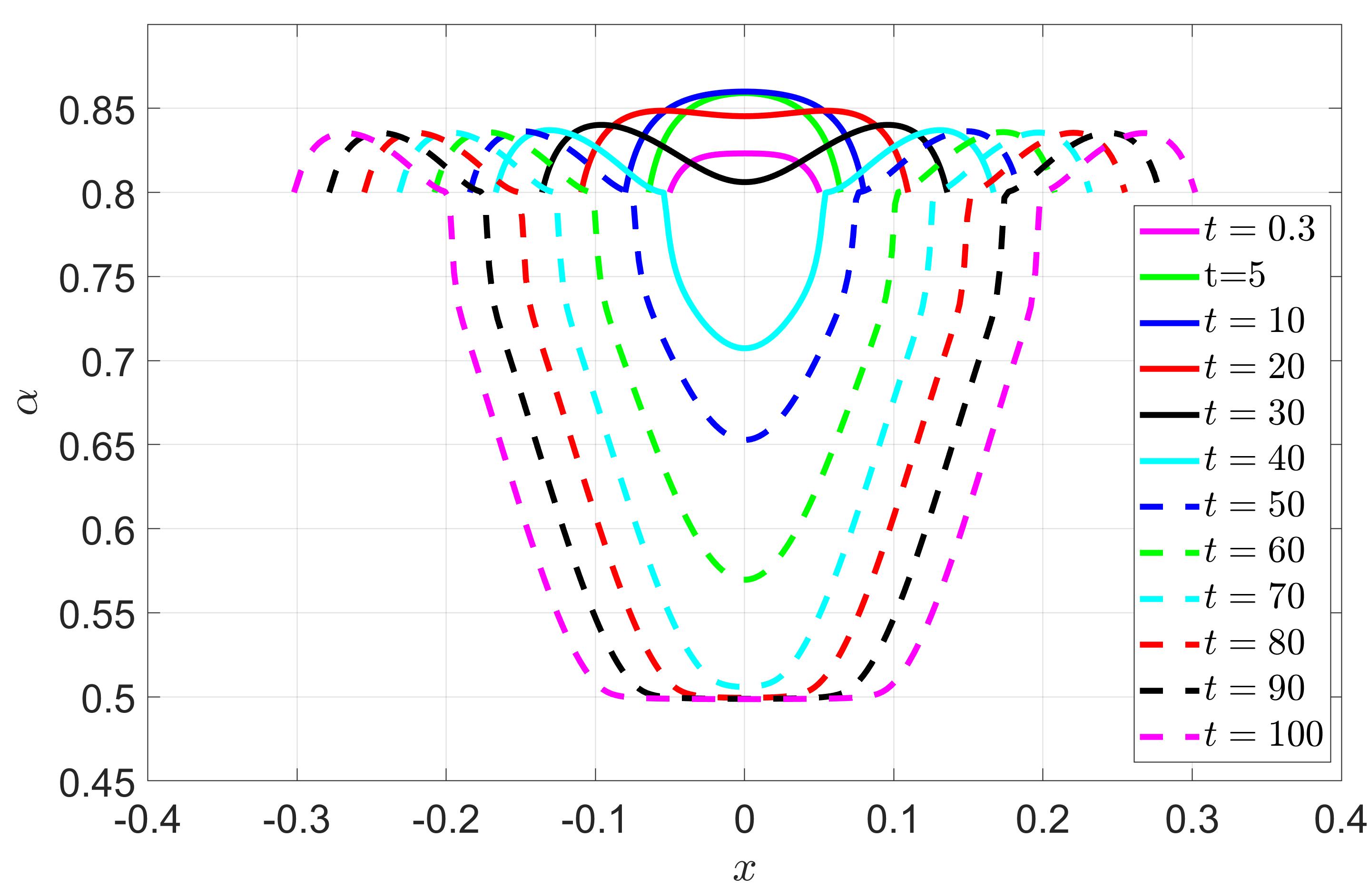}
    		\caption{}
    		\label{fig:volume} 
    	\end{subfigure}
    	\begin{subfigure}{0.49\textwidth}\centering
    		\includegraphics[scale=0.08]{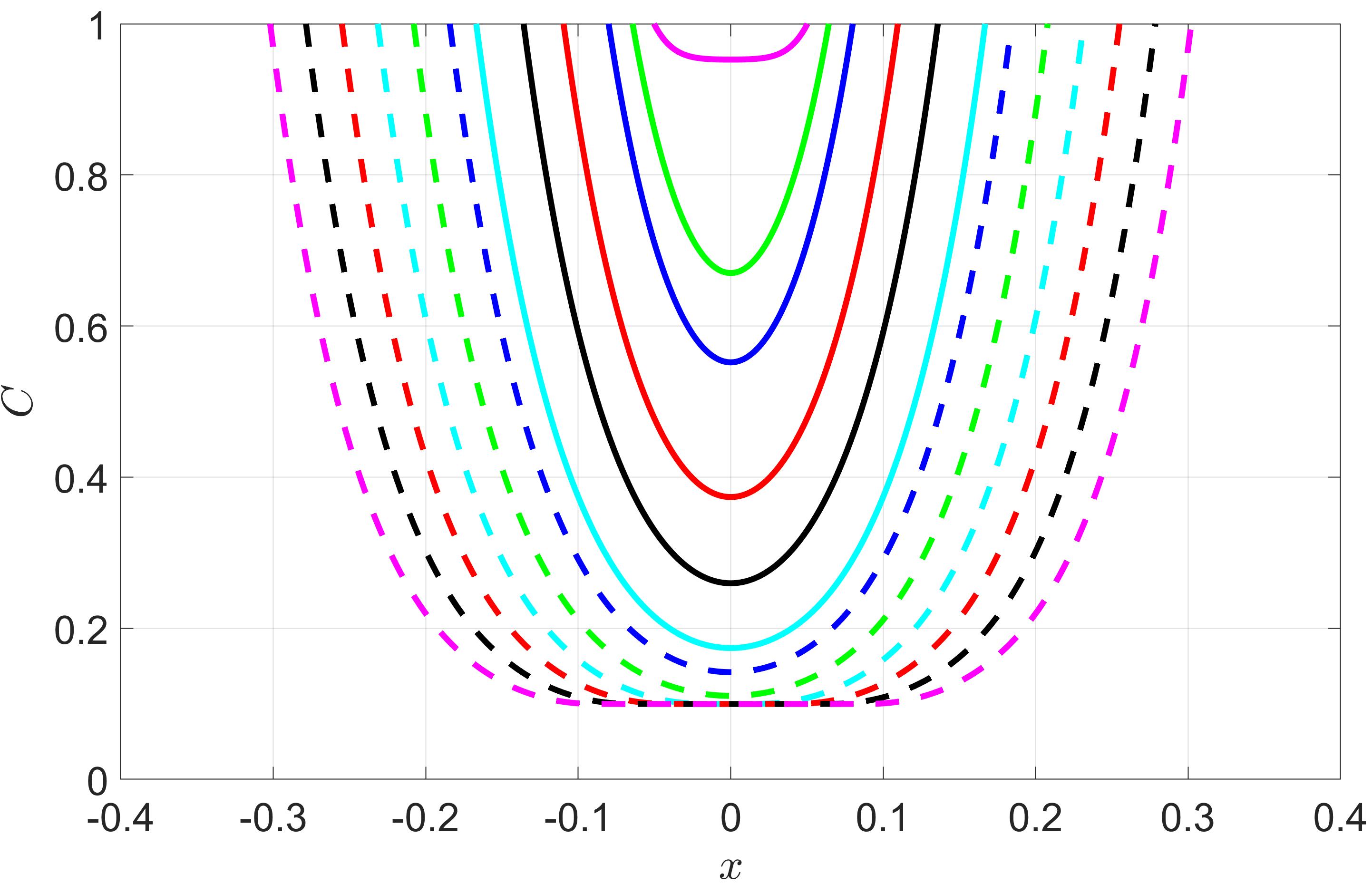}
    		\caption{}
    		\label{fig:oxygen}
    	\end{subfigure} 	
    	\caption[Two numerical solutions]{(a) Volume fraction profile of tumor cellular phase, and (b) oxygen distribution over tumor region ($C_{l}=1$, $C_{r}=1$).}
    	\label{fig:volume_oxygen1}
    \end{figure}

\begin{figure}[h!]
    	\begin{subfigure}{0.49\textwidth}
    		\centering
    		\includegraphics[scale=0.08]{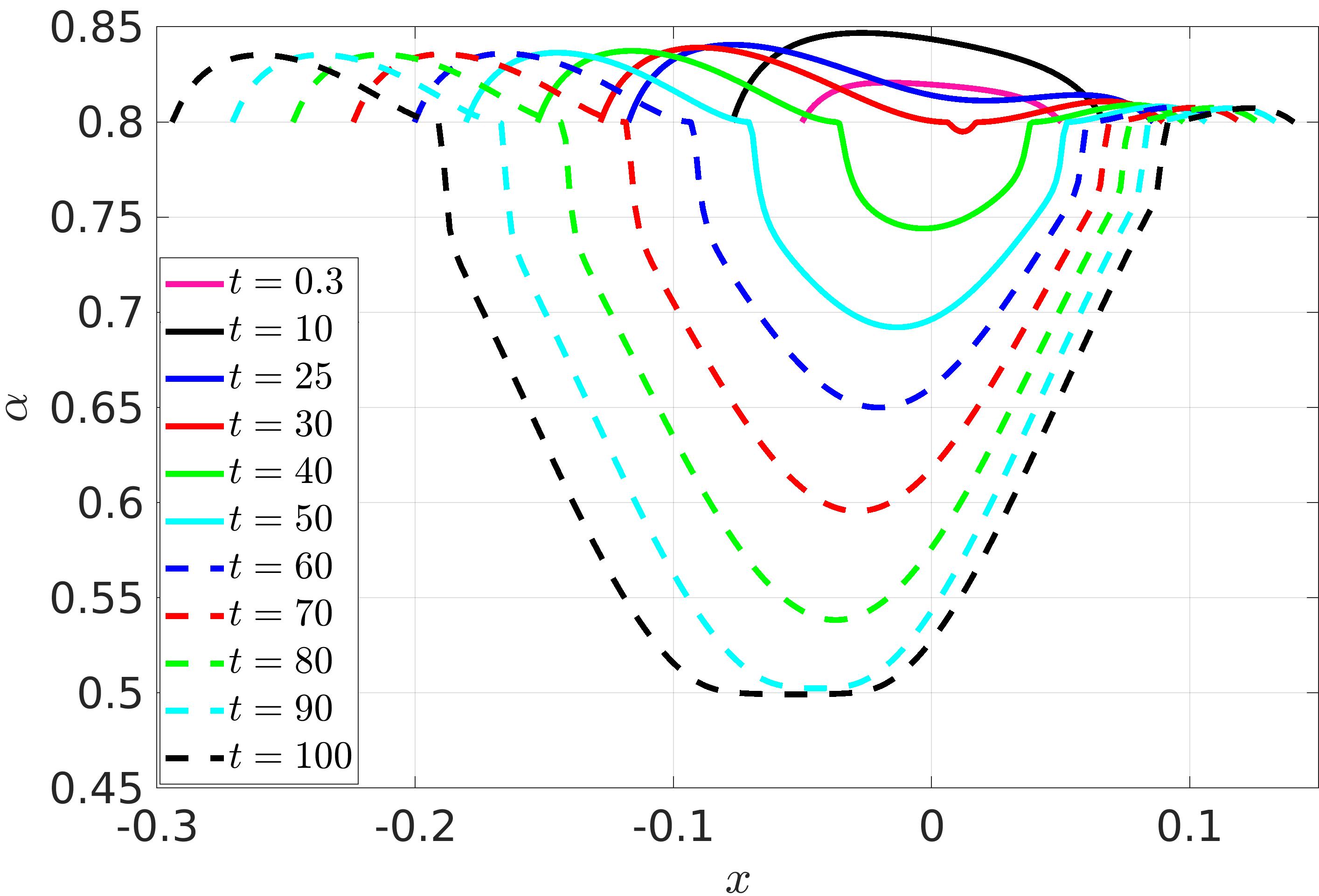}
    		\caption{}
    		\label{fig:volume05} 
    	\end{subfigure}	
    	\begin{subfigure}{0.49\textwidth}
    		\centering
    		\includegraphics[scale=0.08]{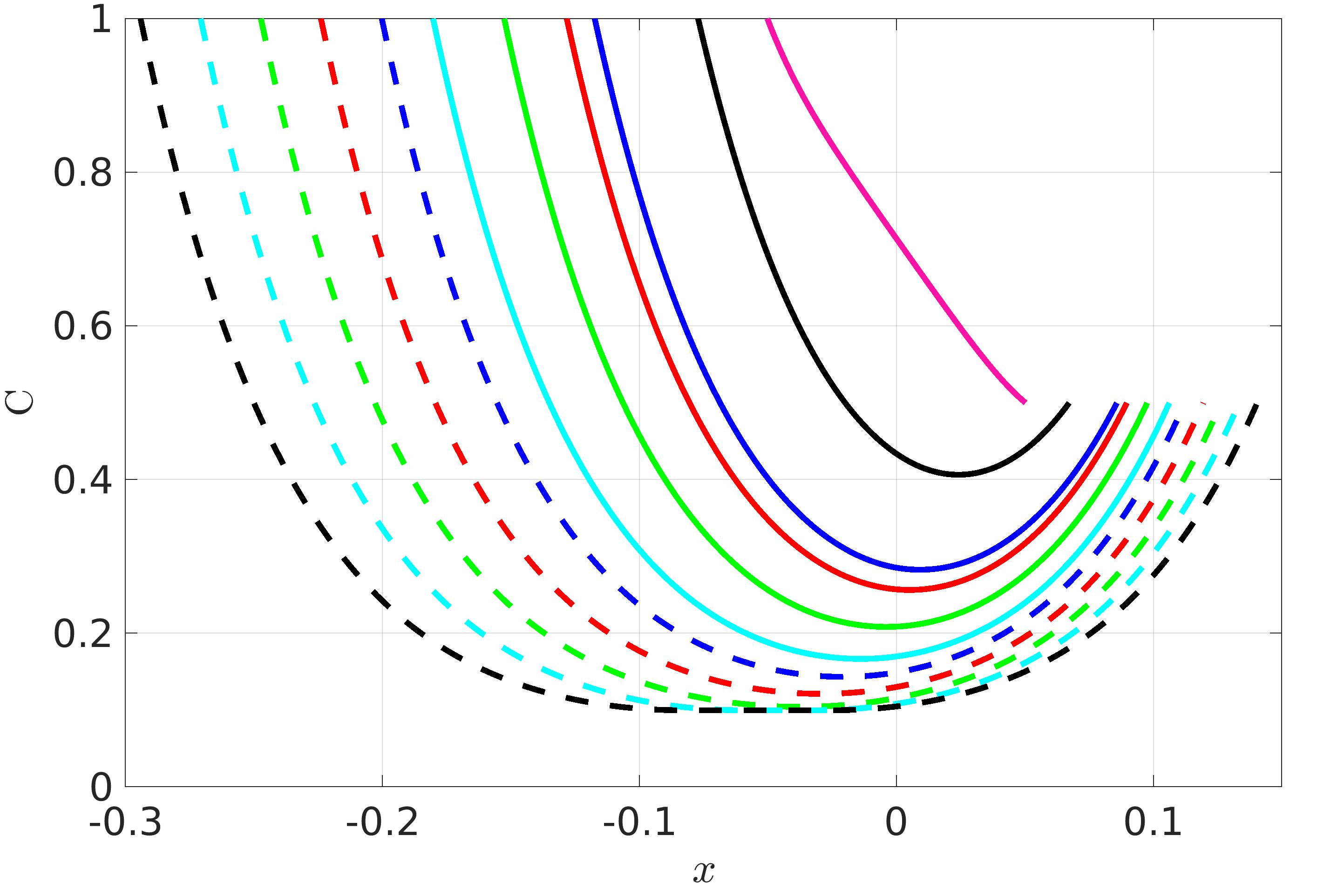}
    		\caption{}
    		\label{fig:oxygen05}
    	\end{subfigure}
    	\caption[Two numerical solutions]{(a) Volume fraction profile of tumor cellular phase, and (b) oxygen distribution over tumor region ($C_{l}=1$, $C_{r}=0.5$).}
    	\label{fig:volume_oxygen2}
    \end{figure}

\begin{figure}[h!]
    	\begin{subfigure}{0.49\textwidth}
    		\centering		
    		\includegraphics[scale=0.08]{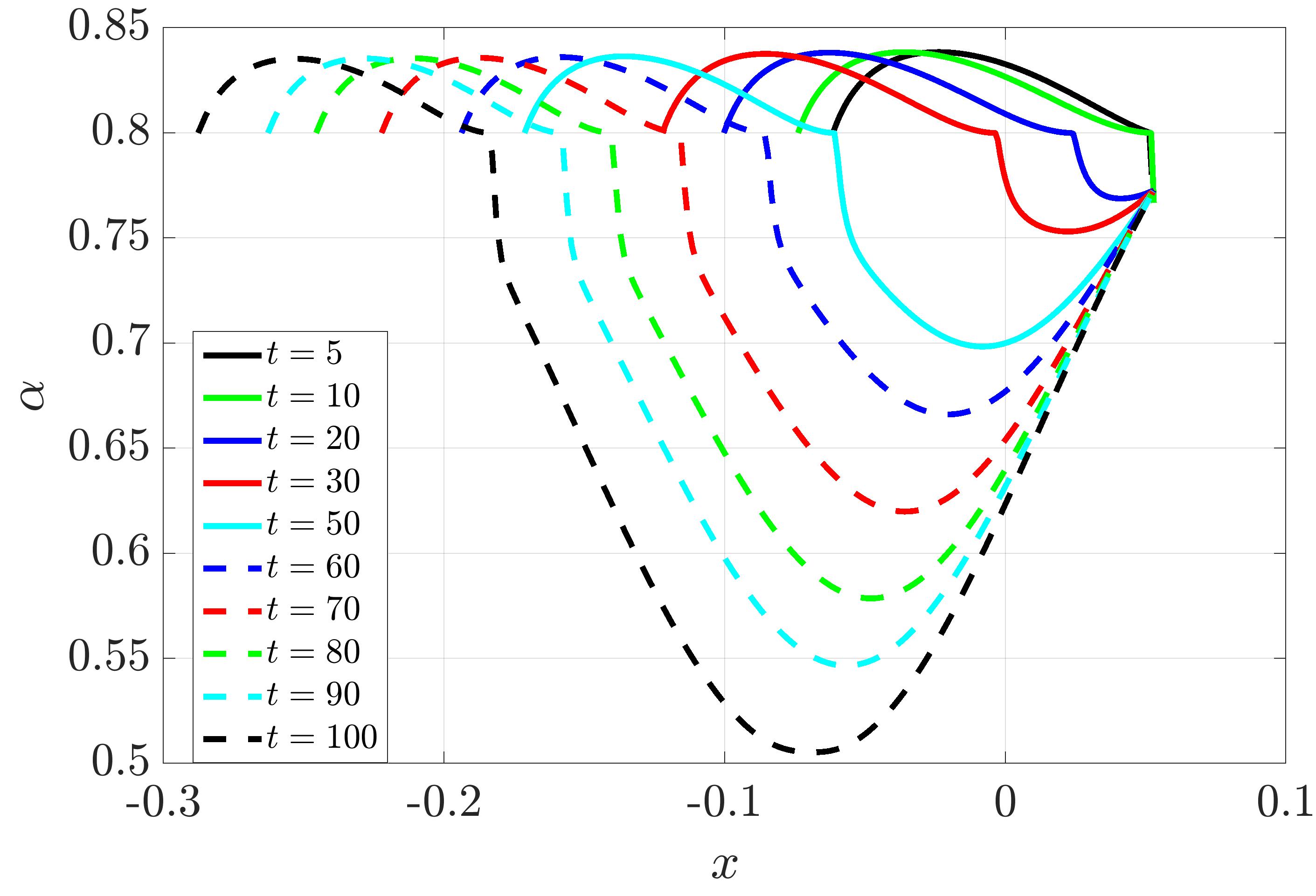}
    		\caption{}
    		\label{fig:volume025} 
    	\end{subfigure}	
    	\begin{subfigure}{0.49\textwidth}
    		\centering		
    		\includegraphics[scale=0.08]{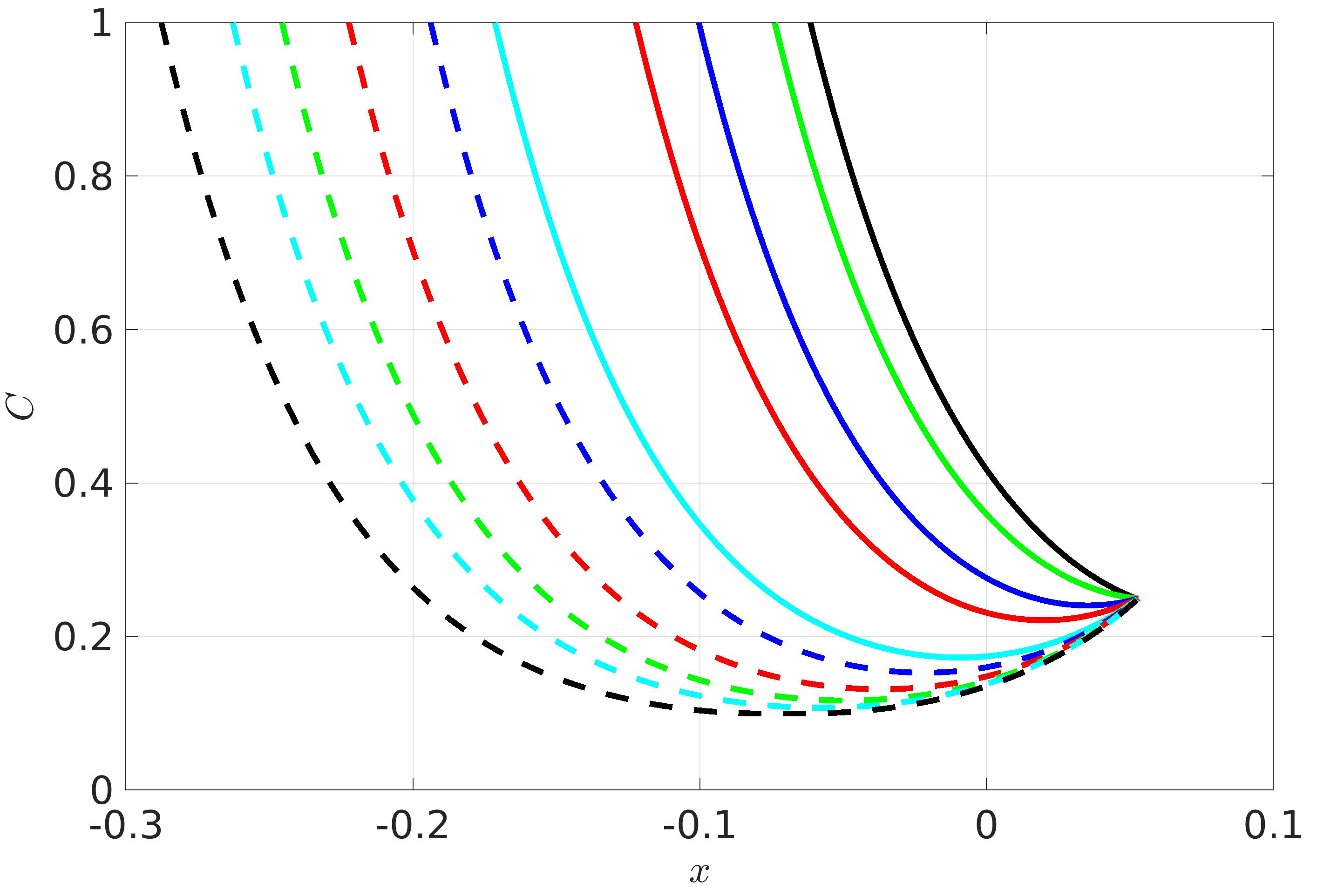}
    		\caption{}
    		\label{fig:oxygen025}
    	\end{subfigure}
    	
    	\caption[Two numerical solutions]{(a) Volume fraction profile of tumor cellular phase, and (b) oxygen distribution over tumor region ($C_{l}=1, C_{r}=0.25$).}
    	\label{fig:volume_oxygen3}
    \end{figure}

\begin{figure}[h!]
    	\begin{subfigure}{0.49\textwidth}
    		\centering
    		\includegraphics[scale=0.08]{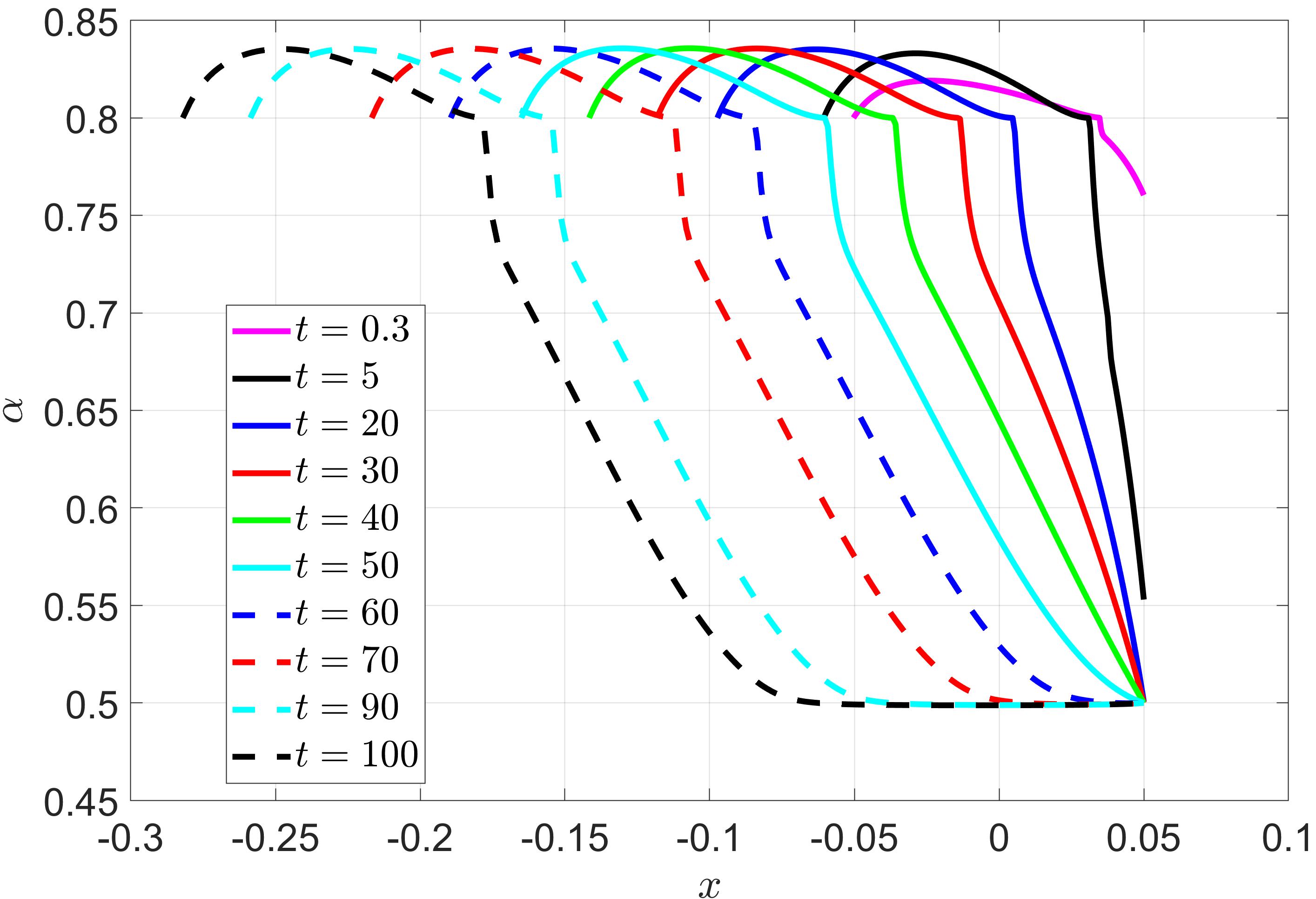}
    		\caption{}
    		\label{fig:volume01} 
    	\end{subfigure}	
    	\begin{subfigure}{0.49\textwidth}
    		\centering
    		\includegraphics[scale=0.08]{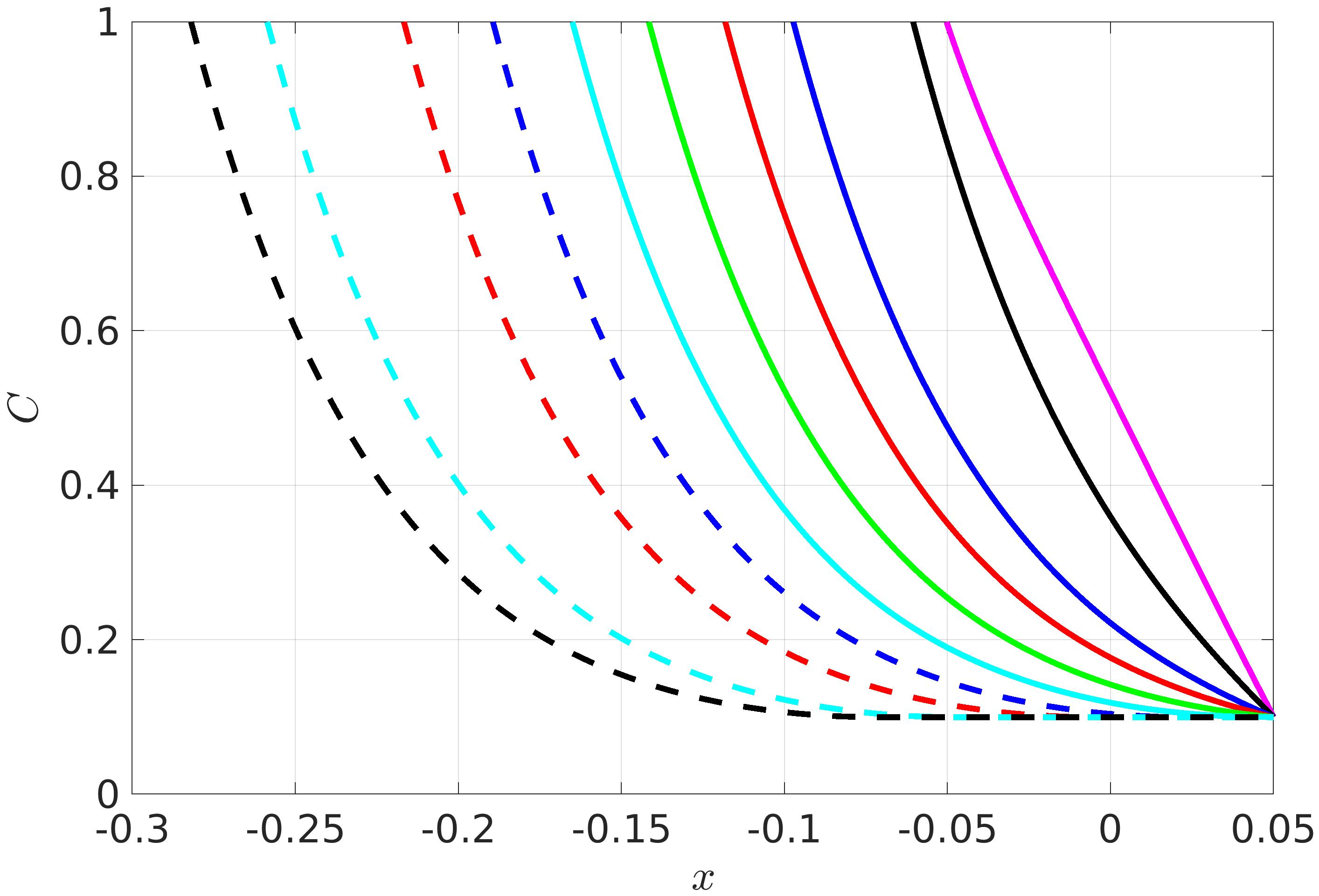}
    		\caption{}
    		\label{fig:oxygen01} 
    	\end{subfigure}
    	\caption{(a) Volume fraction profile of tumor cellular phase, and (b) oxygen distribution over tumor region ($C_{l}=1, C_{r}=0.1$).}
    	\label{fig:volume_oxygen4}
    \end{figure}

    To understand the effects of oxygen supply on tumor cellular volume fraction, again the four cases are considered as $(a)\; C_l=C_r=1$, $(b)\; C_l=1, C_r=0.5$, $(c)\; C_l=1, C_r=0.25$, and $(d)\; C_l=1, C_r=0.1$.
    Changes in the volume fraction of tumor cellular phase and oxygen distribution over tumor region are displayed as subfigures $(a)$ and $(b)$ respectively in each of Figs. \ref{fig:volume_oxygen1}--\ref{fig:volume_oxygen4}.
    
    The volume fraction of cellular phase ($\alpha$) for the first case (i.e., $C_{l}=C_{r}=1$) is displayed in Fig. \ref{fig:volume}. It can be noticed that $\alpha$ increases rapidly with time at initial stage (for $0<t\leq 10$) due to the sufficient supply of oxygen in small tumor. Therefore, the proliferation rate is very high initially. However, as time progresses, though the tumor size increases, the volume fraction of the cellular phase starts falling in the central region of tumor.  
    The oxygen profile for the growing tumor is displayed in Fig. \ref{fig:oxygen}. Clearly, the oxygen concentration falls inside the tumor. As the tumor grows in size, tumor cells die owing to the lack of oxygen supply in the central region. 
  
    As a result, the necrotic region is developed inside the core of the tumor. 
    Also, it can be noticed that the volume fraction of tumor cells is distributed symmetrically about the center $x=0$, and a proliferating rim is spotted near the tumor boundaries.

    In Fig. \ref{fig:volume05}, the cellular volume fraction ($\alpha$) is displayed for oxygen concentrations $C_{l}=1$ and $C_{r}=0.5$. It can be observed that the cells volume is lower near the right boundary as compared to the left one and cells decreases faster at the right boundary. This is because a larger amount of oxygen is available at the left boundary. The distribution of oxygen concentration is displayed in Fig. \ref{fig:oxygen05}. One can observe that the oxygen concentration falls to the threshold value at near $t=80$. The necrotic region starts to develop inside the core, but the shape of this region is asymmetric. Also, $\alpha$ is distributed asymmetrically. Therefore, the tumor morphology is asymmetric.  The findings are quite different from the first case where oxygen concentrations on tumor boundaries are same (Fig. \ref{fig:volume}). One can notice that the layer of proliferating tumor cells is less wide near the right boundary as compared to that of the left boundary. 
    
    Next, the numerical experiments are conducted for $C_l=1$ and $C_r=0.25$. The cellular volume fraction ($\alpha$) is portrayed in Fig. \ref{fig:volume025},whereas the oxygen distribution is displayed in Fig. \ref{fig:oxygen025}. One can see that the tumor growth is negligible towards the right boundary, and $\alpha$ decreases during the initial growth. This is due to the insufficient oxygen supply at the right boundary. As a result, tumor cells start to die there, and the necrotic region is placed in the vicinity of the same boundary. 
    Tumor mostly grows toward the left side due to the sufficient oxygen supply there and the cells proliferating layer also appears only at the left boundary.

    In the next case, the oxygen concentration is considered to be the same as the necrotic threshold value at the right boundary (i.e., $C_r=C_N=0.1$). The oxygen profile for the growing tumor is portrayed in Fig. \ref{fig:oxygen01}, and the volume fraction of cellular phase ($\alpha$) is shown in Fig. \ref{fig:volume01}. One can see that $\alpha$ starts declining near the right boundary as time progresses. It achieves the lowest value $\alpha=0.5$ at the time $t=20$. It can be noticed that the tumor growth is negligible towards the right boundary as a result of the fact that the oxygen is insufficient to sustain the tumor cells. So, tumor cells are unable to proliferate owing to the tumor cells death.
    
\begin{figure}[h!]\centering
    	\begin{subfigure}{0.49\textwidth}
    		\centering
    		\includegraphics[scale=0.08]{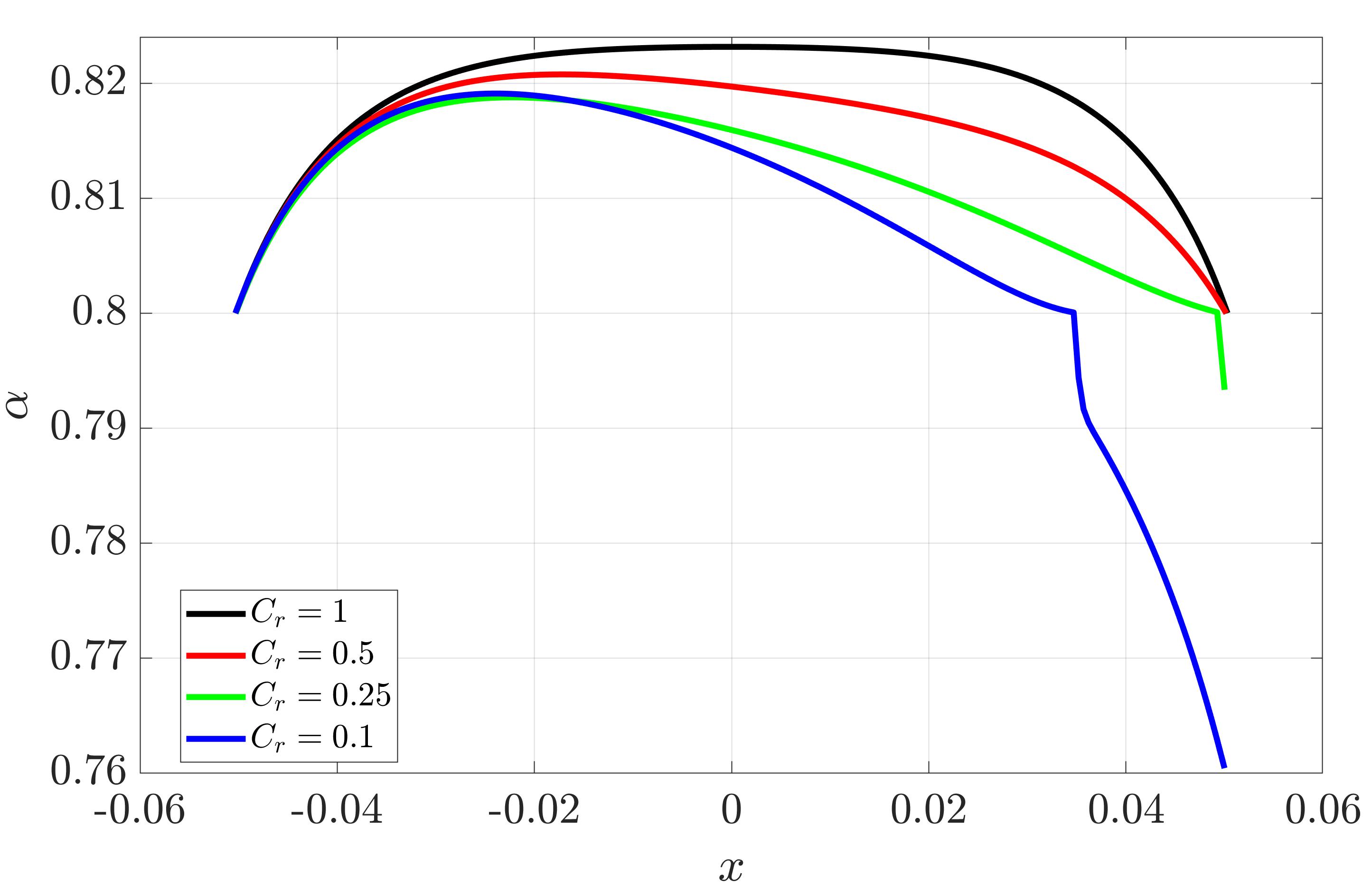}
    		\caption{}
    		\label{fig:tt0.3}
    	\end{subfigure}	
    	\begin{subfigure}{0.49\textwidth}
    		\centering
    		\includegraphics[scale=0.08]{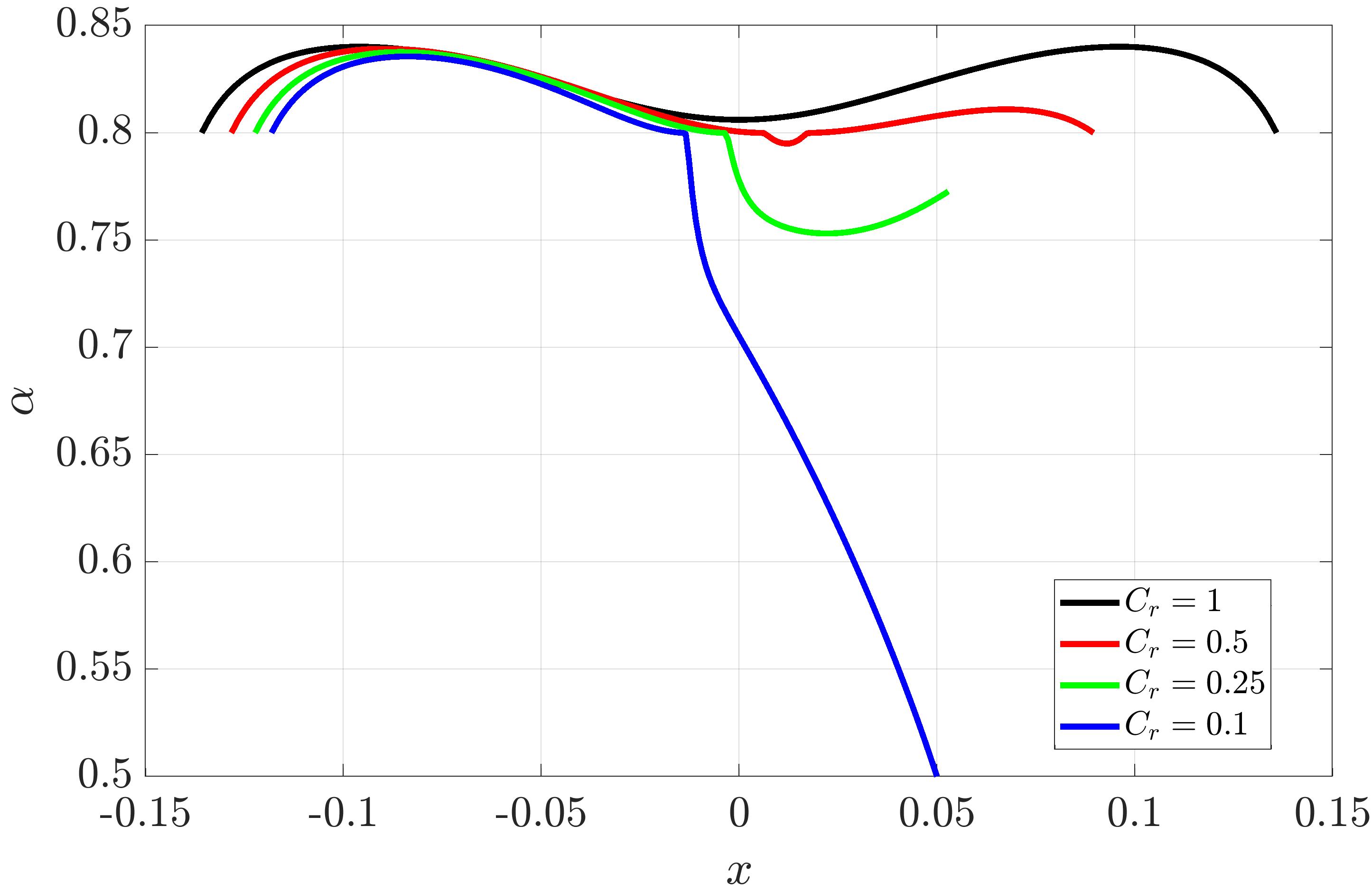}
    		\caption{}		
    		\label{fig:tt30}		
    	\end{subfigure}		
    	\begin{subfigure}{0.49\textwidth}
    		\centering
    		\includegraphics[scale=0.08]{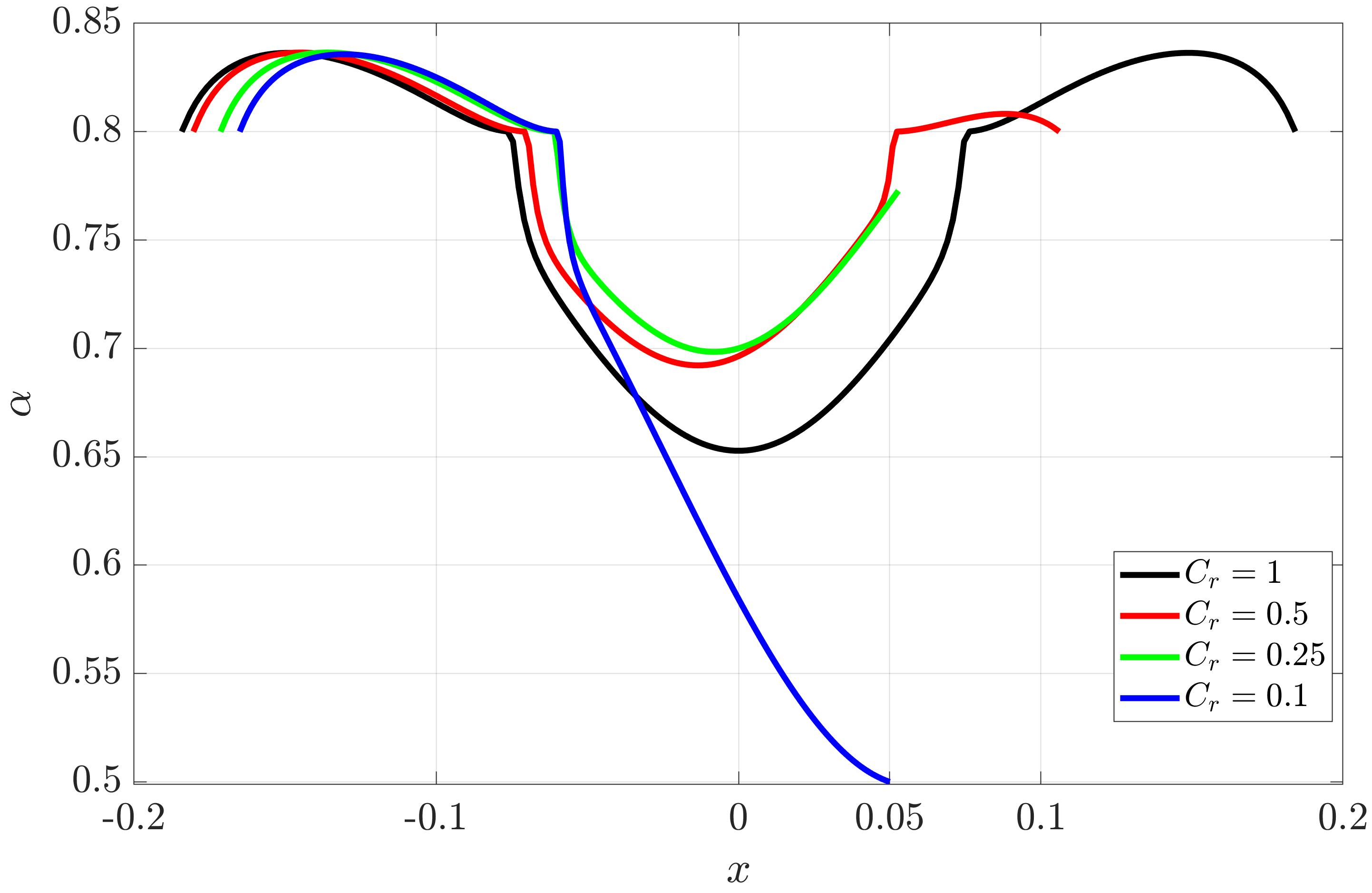}
    		\caption{}		
    		\label{fig:tt50}
    	\end{subfigure}
    	\begin{subfigure}{0.49\textwidth}
    		\centering
    		\includegraphics[scale=0.08]{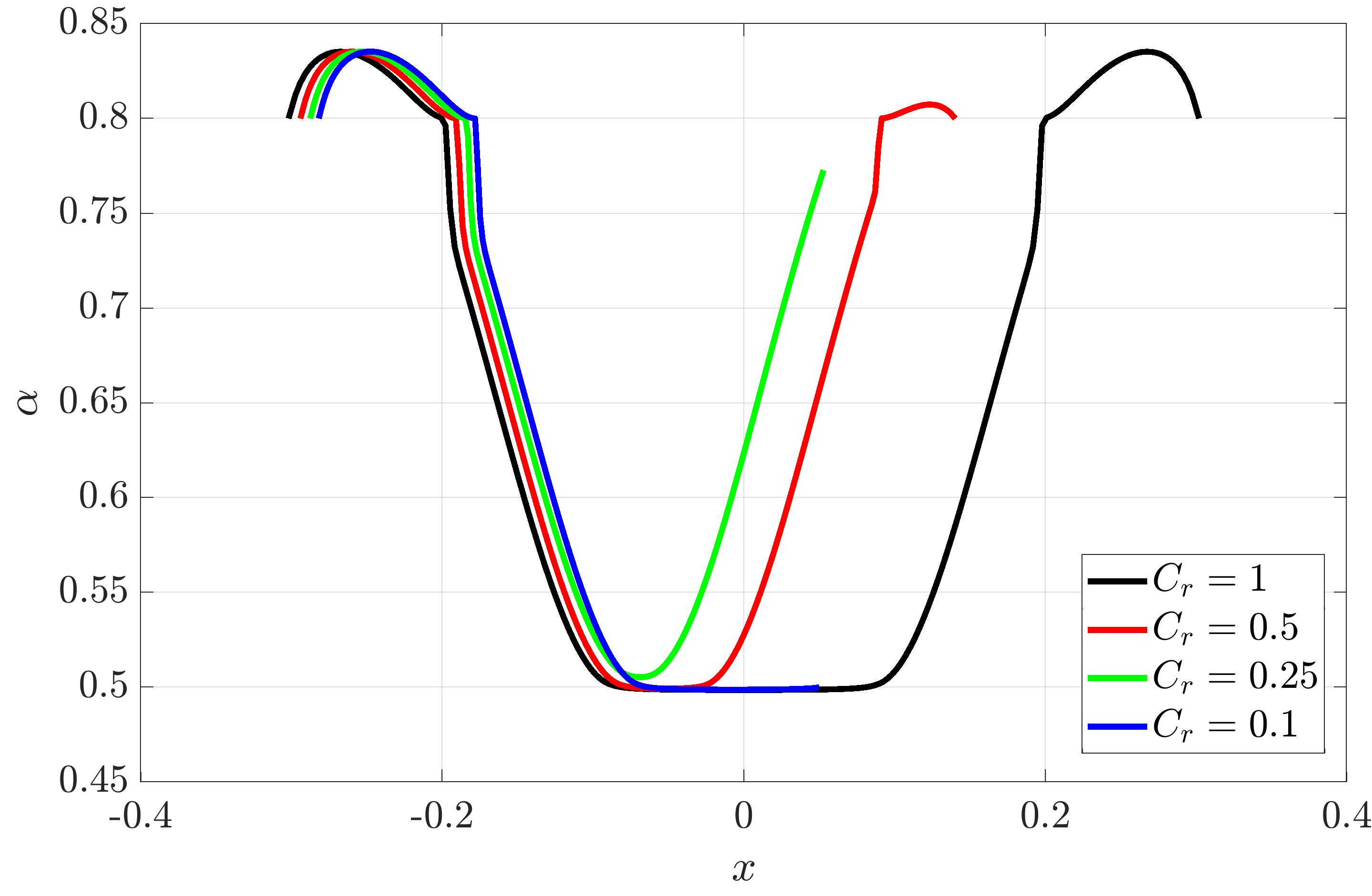}
    		\caption{}		
    		\label{fig:tt100}
    	\end{subfigure}
    	\caption{Comparison between volume fraction of tumor cellular phase for different oxygen concentration at right boundary at (a) $t=0.3$, (b) $t=30$, (c) $t=50$, and (d) $t=100$.}
    	\label{fig:volume_compr}
    \end{figure}

  Comparison between the volume fraction of tumor cellular phase for different oxygen concentrations at the right boundary is portrayed at Fig. \ref{fig:volume_compr}. The snapshots Figs. \ref{fig:tt0.3}, \ref{fig:tt30}, \ref{fig:tt50}, and \ref{fig:tt100} are taken at $t=0.3, 30, 50, \text{ and } 100$, respectively. It can be noticed that tumor volume fraction increases through-out the domain for the cases $C_r=1$ and $C_r=0.5$ at $t=0.3$. On the other hand, tumor cells start to die at the boundary having oxygen concentration $C_r=0.25$ and $C_r=0.1$. This is due to oxygen supply that is not sufficient to fulfill to the demand for livelihood of cells (Fig. \ref{fig:tt0.3}). However, at subsequent times ($t=30, 50, \text{ and } 100$), tumor cells also die in the central region for each combination of oxygen concentration at the right boundary. One can observed that the proliferation rim is spotted at the right edge of the tumor only for $C_r=1$ and $C_r=0.5$ (Figs. \ref{fig:tt30}--\ref{fig:tt100}).
 It can be concluded that tumor volume fraction decreases at the right boundary with the decreasing value of oxygen concentration at that boundary, and the shape of tumor strongly depends on the oxygen supply. 
   
	\subsection{Effects of oxygen supply on cellular phase velocity}
	The cellular phase velocity determines the aggressive nature of the tumor and its progression. So, its understanding is vital for developing new therapies for cancer disease.
	In this subsection, the effects of oxygen on tumor cellular phase velocity is investigated.

\begin{figure}[h!]\centering
		\begin{subfigure}{0.495\textwidth}
			\centering
			\includegraphics[scale=0.08]{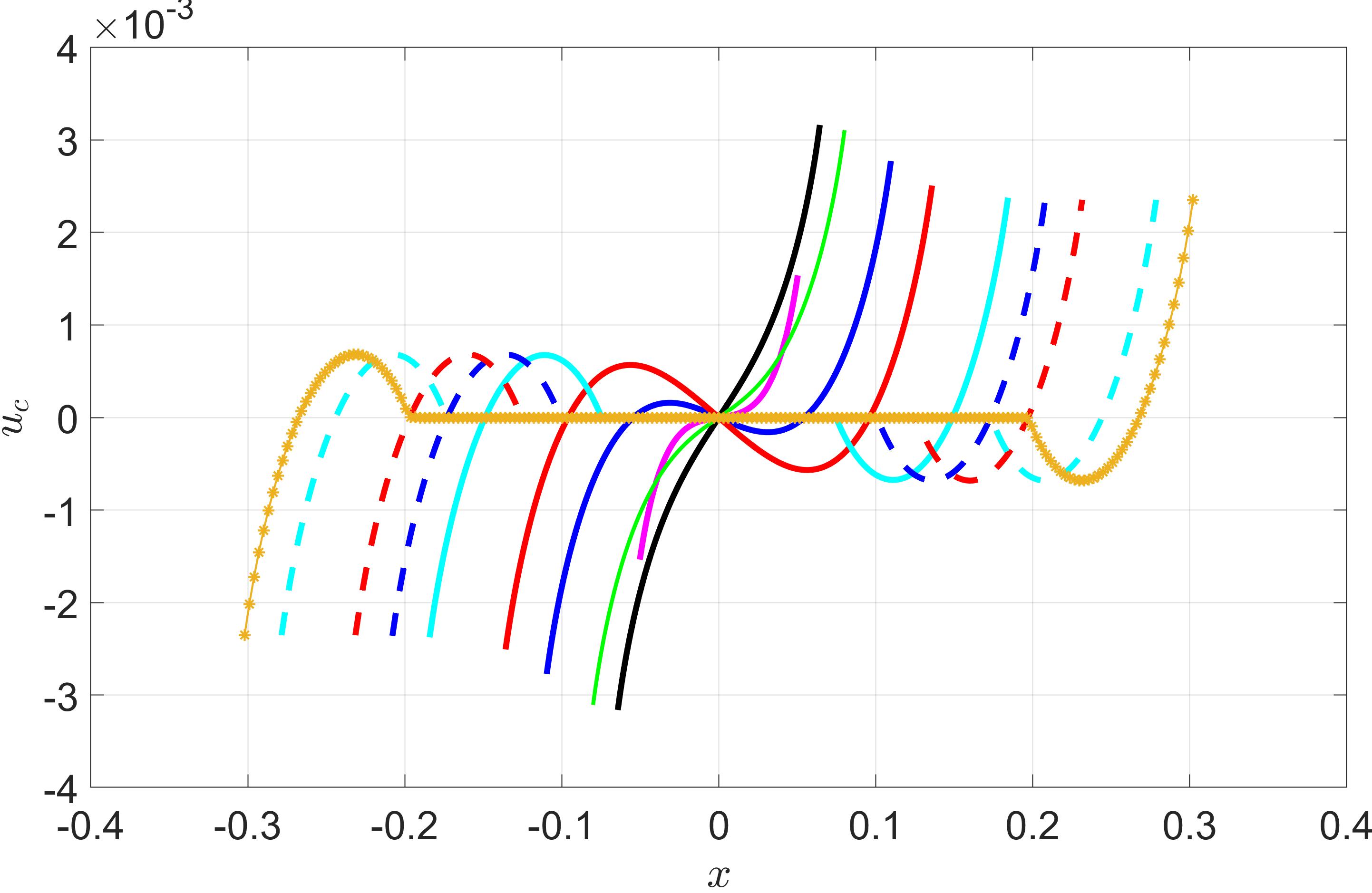}
			\caption{}
			\label{fig:velocity}
		\end{subfigure}	
		\begin{subfigure}{0.495\textwidth}
			\centering
			\includegraphics[scale=0.08]{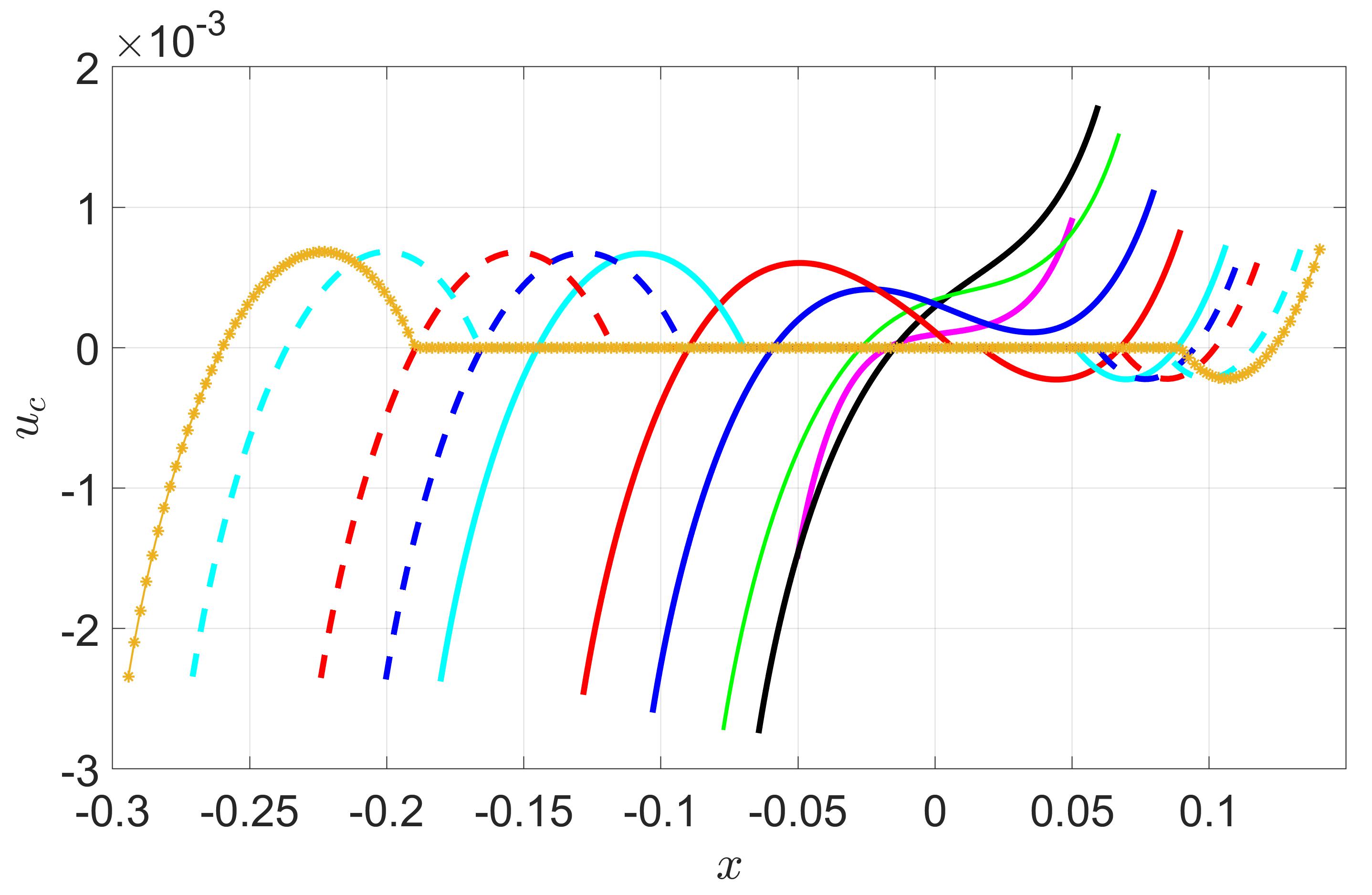}
			\caption{}		
			\label{fig:velocity05}		
		\end{subfigure}		
		\begin{subfigure}{0.495\textwidth}
			\centering
			\includegraphics[scale=0.08]{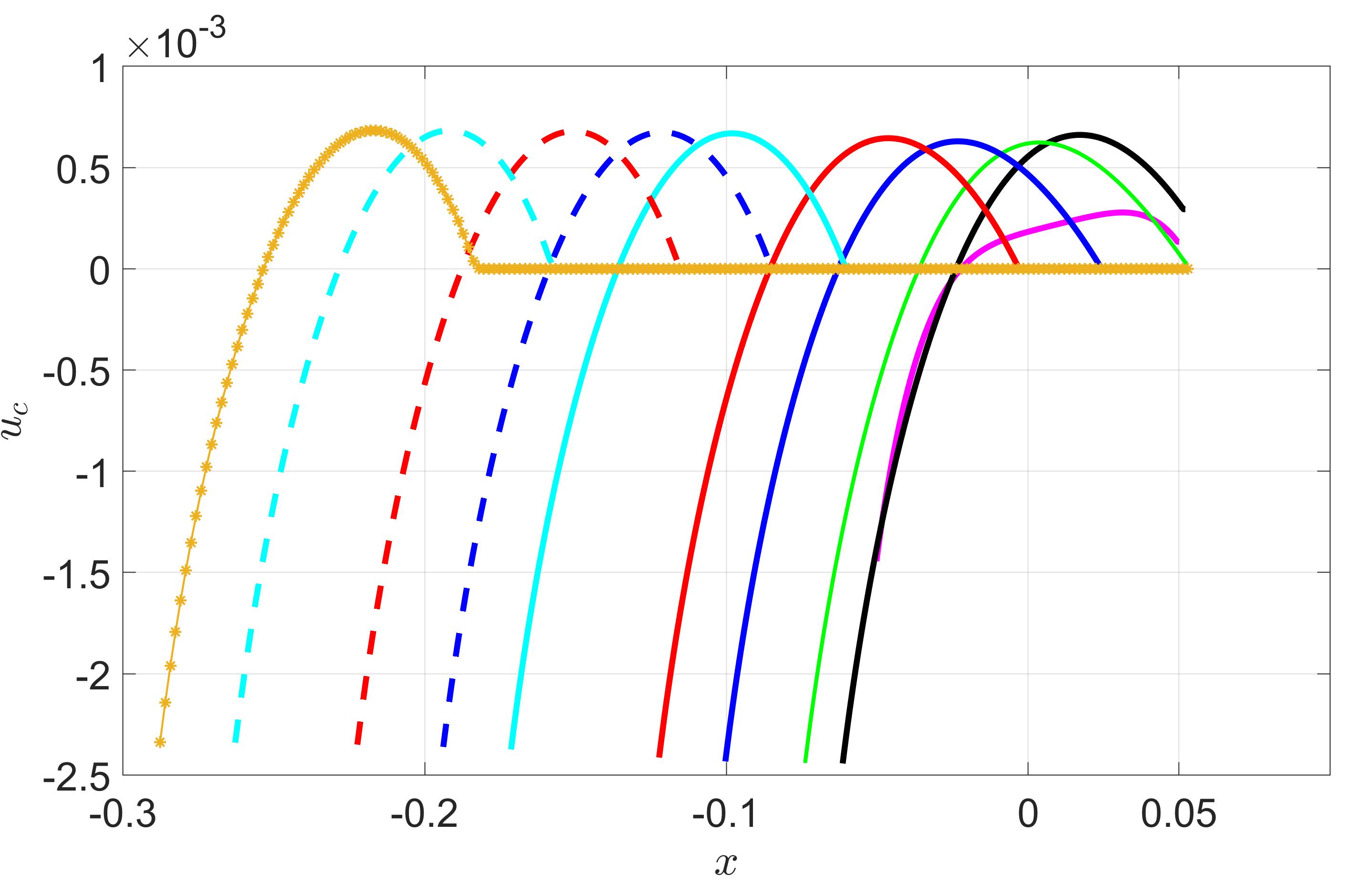}
			\caption{}		
			\label{fig:velocity025}
		\end{subfigure}
		\begin{subfigure}{0.495\textwidth}
			\centering
			\includegraphics[scale=0.08]{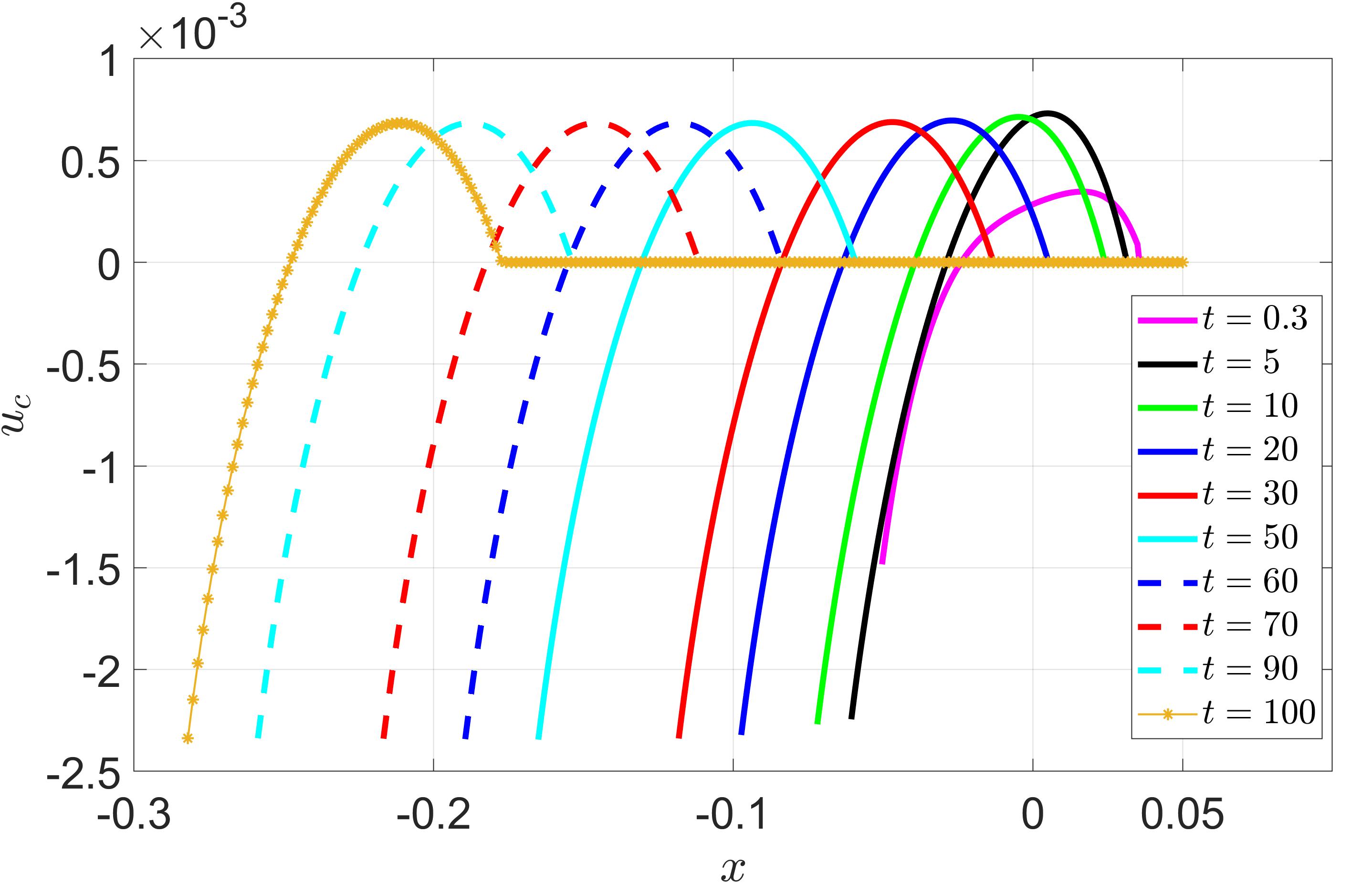}
			\caption{}		
			\label{fig:velocity01}
		\end{subfigure}
		\caption{Velocity of tumor cellular phase over the tumor region with oxygen boundary concentrations (a) $C_l=C_r=1$, (b) $C_l=1$,  $C_r=0.5$, (c) $C_l=1$, $C_r=0.25$, and (d) $C_l=1$, $C_r=0.1$.}
	\end{figure}

   The velocity profiles of the cellular phase with boundary oxygen concentrations $C_l=1, C_r=1$ are plotted at different times in Fig. \ref{fig:velocity}. The velocity is anti-symmetric about $x=0$, and it increases with $x$ during the initial stage of the development of tumor ($0< t \leq 5$). 
	However, as time progresses, the velocity decreases mainly in the central region of the tumor. In the time interval $5<t\le 30$, on the right side of $x=0$, the velocity changes from negative to positive. While the velocity changes from positive to negative as we move to the left from $x=0$.
	This is due to the dead cells, which are converted into the ECM during tumor progression. For $t>30$, the velocity becomes zero at the central region of the tumor and this region is termed as necrotic core of the tumor, which grows in size as tumor progresses. Owing to the sufficient oxygen supply, the tumor spreads through its boundaries, and it grows symmetrically about $x=0$ with the same magnitude of velocity on both sides of the tumor.

	For the second case, where $C_l=1$ and $C_r=0.5$, the velocity profile of the cellular phase is portrayed in Fig. \ref{fig:velocity05}. The magnitude of the velocity of cellular phase at the left boundary is approximately 1.5 times the velocity at the right boundary for $t=0.3$. The velocity increases continuously on both sides of the boundary up to $t=10$. It shows the rapid proliferation of tumor cells initially. However, as time progresses, cellular velocity decreases until $t=50$ and becomes stagnant $t=60$ onward. The magnitude of the velocity of cellular phase at the left boundary is higher than that at the right boundary. 
	
	The velocity profile  for the case $C_{r}=0.25$ and $C_l=1$ is displayed in Fig. \ref{fig:velocity025}. In this case, a very small progression of the right boundary is spotted during the initial stage of tumor growth (i.e., for $0<t\leq5$). The velocity of the tumor cellular phase becomes zero at the right boundary $t=20$ onward. whereas, the cellular phase velocity at the left boundary increases over time for $0<t\leq10$ and achieves a steady-state for $t\geq10$. 
	 \begin{figure}[h!]\centering
	 	\begin{subfigure}{0.495\textwidth}
	 		\centering
	 		\includegraphics[scale=0.08]{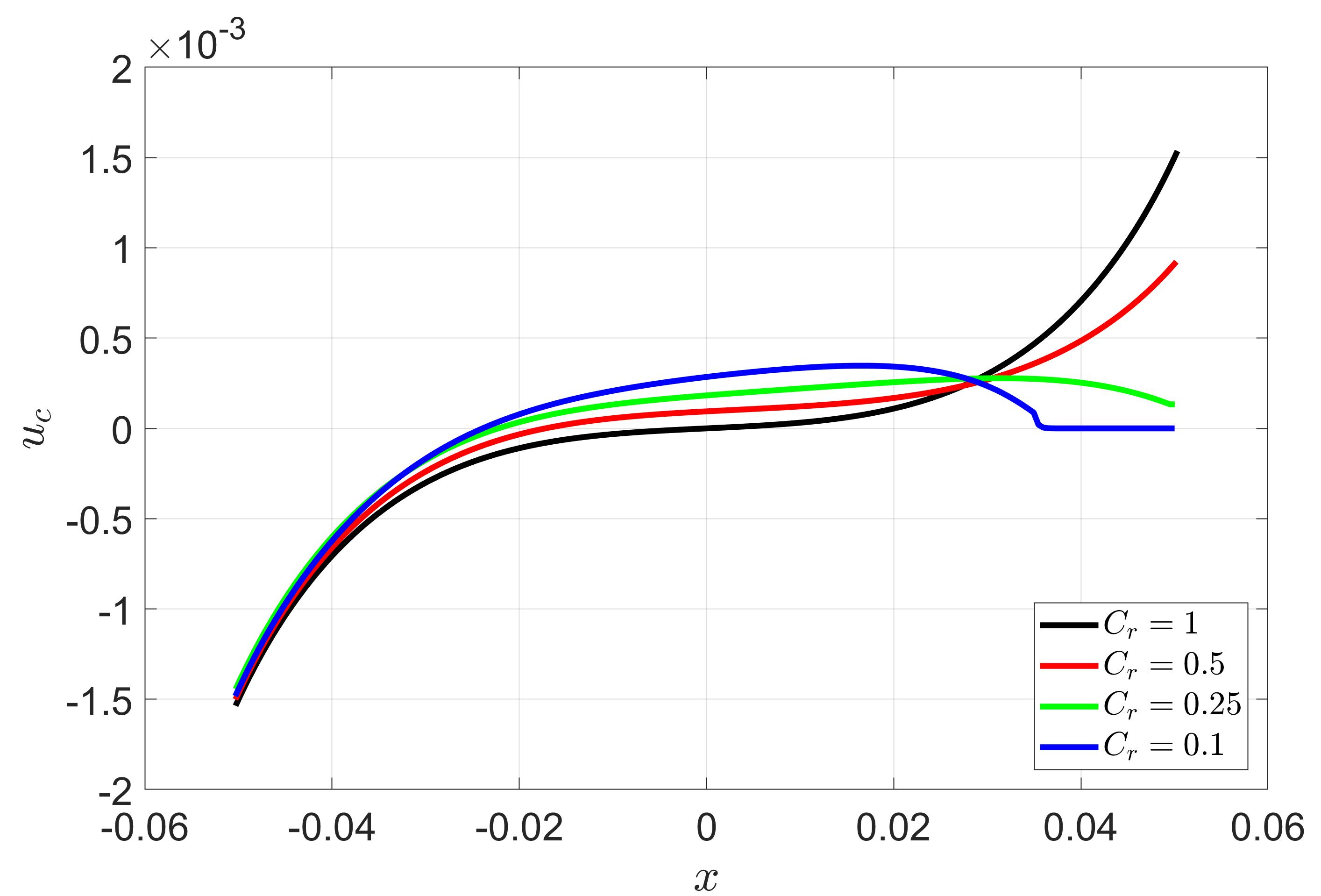}
	 		\caption{}
	 		\label{fig:t0.3}
	 	\end{subfigure}	
	 	\begin{subfigure}{0.495\textwidth}
	 		\centering
	 		\includegraphics[scale=0.08]{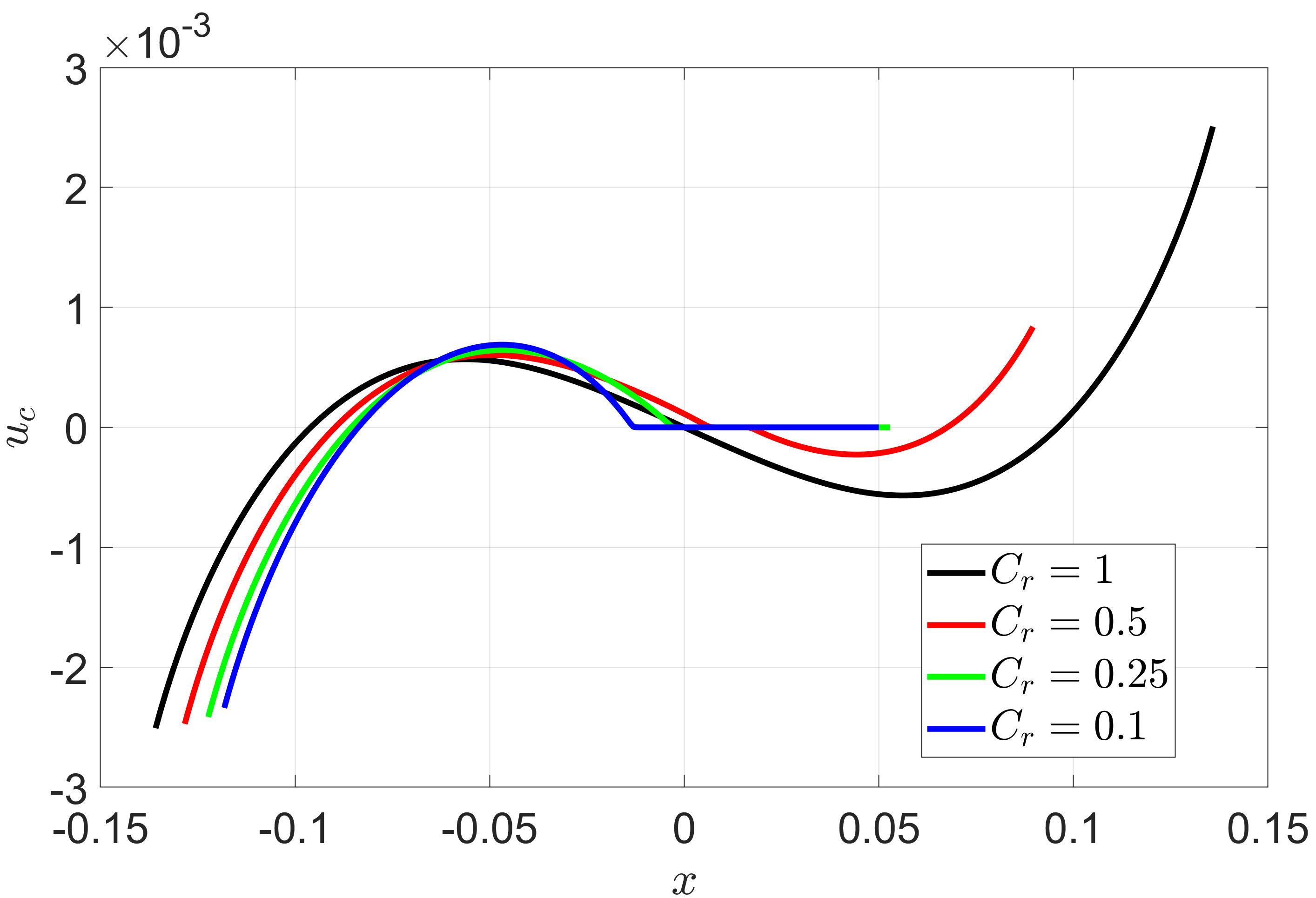}
	 		\caption{}		
	 		\label{fig:t30}		
	 	\end{subfigure}		
	 	\begin{subfigure}{0.495\textwidth}
	 		\centering
	 		\includegraphics[scale=0.08]{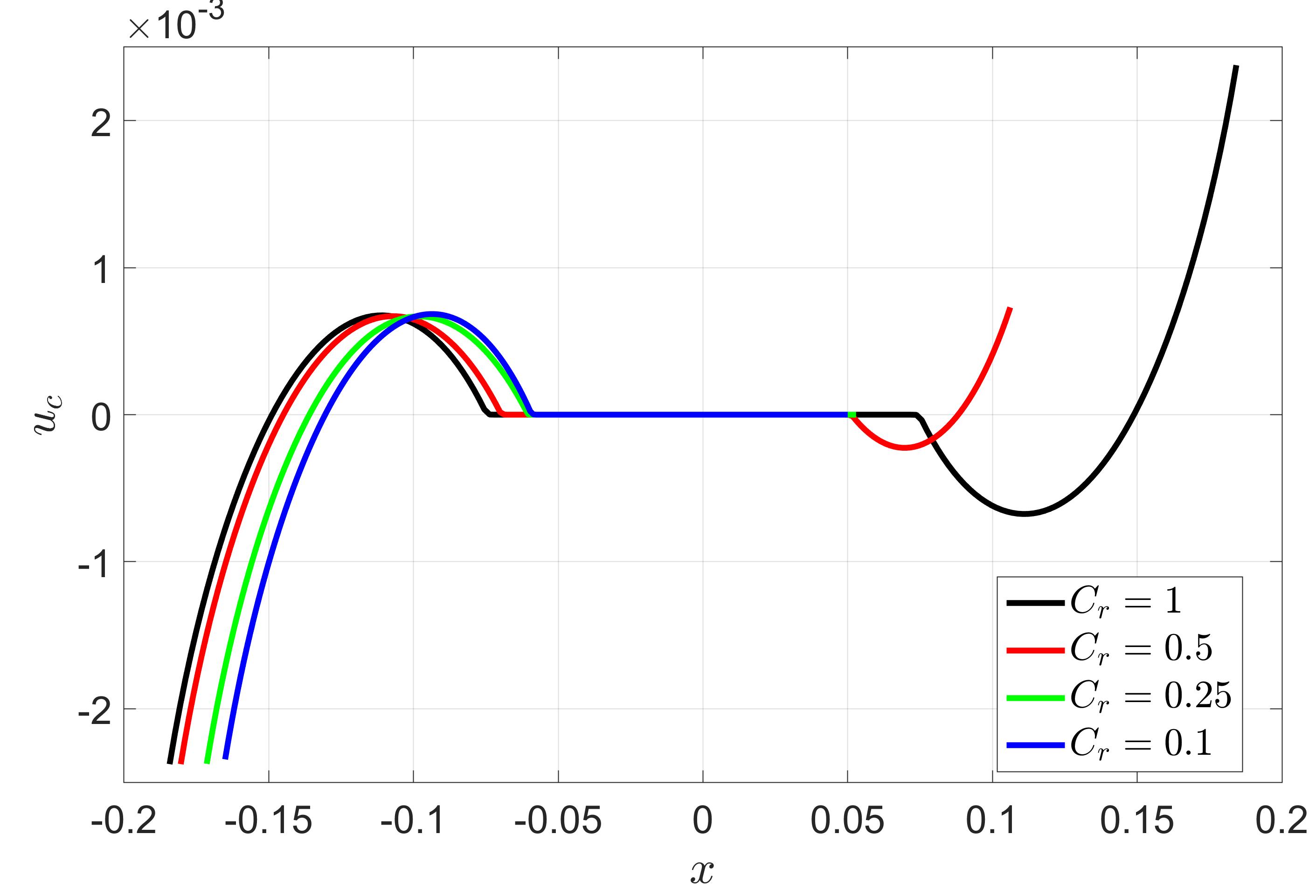}
	 		\caption{}		
	 		\label{fig:t50}
	 	\end{subfigure}
	 	\begin{subfigure}{0.495\textwidth}
	 		\centering
	 		\includegraphics[scale=0.08]{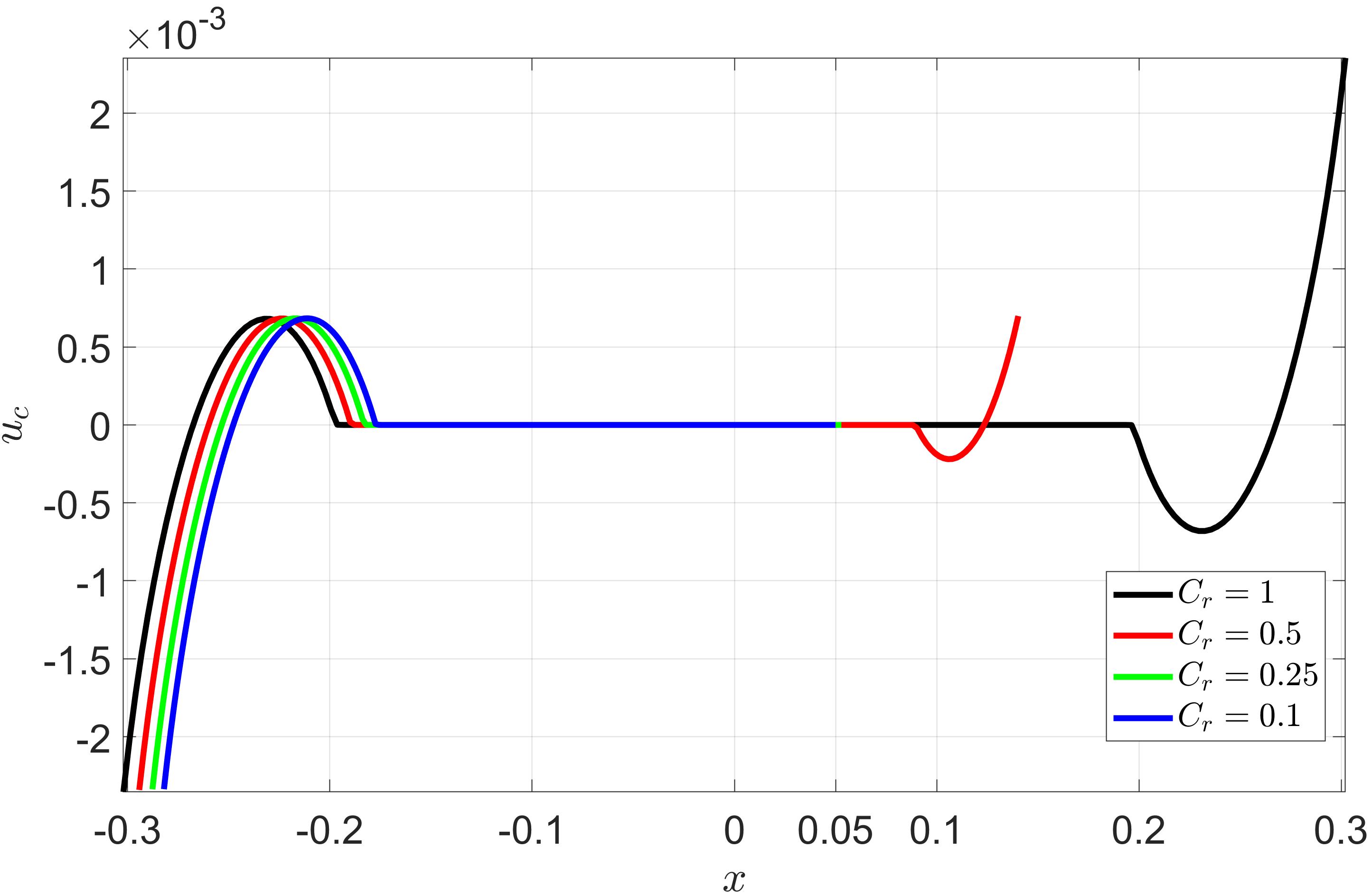}
	 		\caption{}		
	 		\label{fig:t100}
	 	\end{subfigure}
	 	\caption{Comparison between velocity profiles of tumor cellular phase for different oxygen concentration at right boundary at (a) $t=0.3$, (b) $t=30$, (c) $t=50$, and (d) $t=100$.}
	 	\label{fig:velocityCompare}
	 \end{figure}
	
The case with $C_{l}=1$ and $C_{r}=C_N=0.1$ reflects the growth dynamics under poor supply of oxygen at the right boundary. This situation mimics in-vivo situation when some part of the tumor is far from the blood vessels. The velocity profile of cellular phase over the time is displayed in Fig. \ref{fig:velocity01}. There is no movement of tumor cellular phase near the right boundary. This is owing to the oxygen supply that remains below the threshold value of necrotic cell death. Whereas, the velocity near the left boundary increases over time up to $t=30$ and achieves steady-state when $t\geq30$. So, it can be concluded that the tumor grows under the sufficient oxygen supply, and the tumor growth could be significant near the blood vessels.

Fig. \ref{fig:velocityCompare} shows the velocity profiles of the cellular phase for different oxygen supplies at the right edge of tumor at different times. 
		It can be observed that the magnitude of the cellular phase velocity at the right boundary is always higher with $C_r=1$ in comparison to the other cases where $C_r=0.5,0.25,0.1$. In Fig. \ref{fig:t0.3}, one can find that the velocity at the left boundary has a negligible effect on its magnitude while varying the oxygen concentration at the other end of the tumor. 
		It manifests that having unequal oxygen supplies at one boundary do not affect the other boundary propagation characteristics as far as the growth is concerned with the oxygen only. This  might be due to the large necrotic core that does not allow the exchange of the proliferating cells between the two sides.

	\subsection{Effects of oxygen supply on diameter}
	The diameter of a tumor could be a good measurement of the tumor size and can help to identify the rate of its growth. The size of a tumor is used as a tool to recognize the stage of a cancer, which is critical for the choice of a treatment and predicting the patient's prognosis. In this subsection, the effects of oxygen supply on tumor size is investigated. The diameter is measured as the distance of the position of the right boundary to the left boundary.
	
\begin{figure}[h!]\centering
		\includegraphics[scale=0.1]{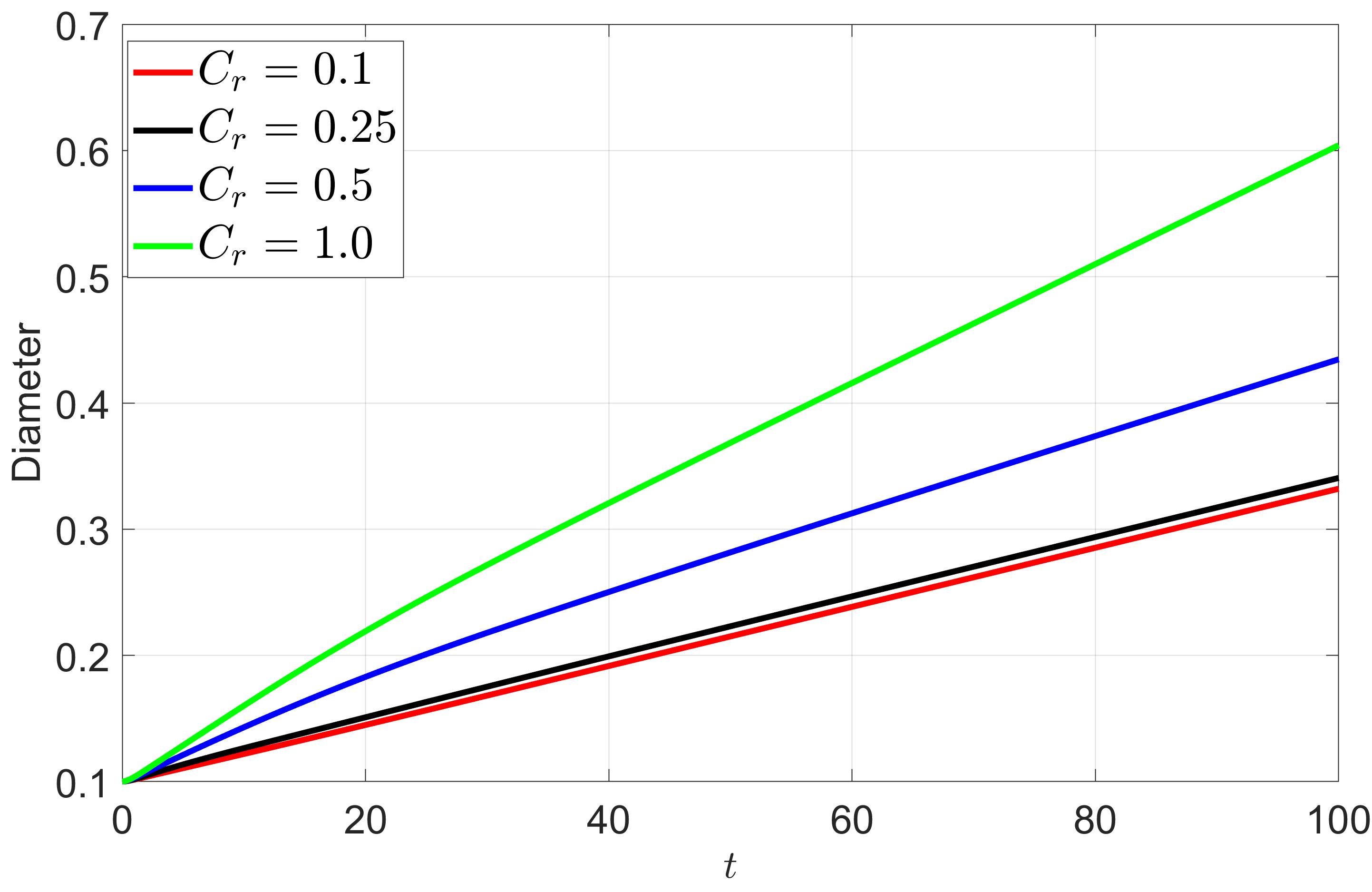}
		\caption{Propagation of diameter of a tumor with different oxygen supplies at right boundary.}
		\label{fig:diametre}
	\end{figure}

	The results on the effects of oxygen supply on the tumor diameter over time are shown in Fig. \ref{fig:diametre}. The oxygen concentration  varies  at the right boundary but is kept fixed ($C_l=1$) at the left boundary.  It can be seen that the diameter of the tumor increases with the increase in oxygen concentration at the right boundary. It can be concluded that the diameter of the tumor strongly depends on the oxygen supply through the surrounding medium. Tumor diameter is larger for the tumor having equal oxygen concentration at the boundaries compared to that for other choices of oxygen concentration at the right boundary.
	
\section{Discussion}
In this work, a spatio-temporal mathematical model is developed to investigate the asymmetric growth of the solid tumor. The effects of tumor boundary oxygen concentrations on tumor growth are analyzed. In the process of avascular tumor growth, nutrients are consumed by living cells. As the tumor grows, the supply of nutrients gradually declines toward the tumor center. So, cells start to die due to starvation in the central region of tumor. These results lead to the formation of a necrotic zone, which increases in size as the tumor grows. This phenomenon has been found to impact the growth dynamics of metastasis tumor \cite{lewin2020three,2016metastsis_hypoxic}, which has been a leading cause of death due to cancer \cite{2019stats}. So, analysis of tumor growth is crucial for development of better treatments and prevention of metastasis processes. In reality, a tumor is a mixture of various malignant cells, healthy cells, ECM, and immune cells \cite{2007Li}; however, the present model incorporates the tumor cells and ECM only. The immune cells are ignored owing to the compromised immune system \cite{2008PRESTWICH}. 
	
Tumor cells seek for energy to proliferate and maintain their livelihood. The energy in tumor cells is generated from many sources \cite{hanahan2000hallmarks}. Glucose, lactate, and oxygen play a significant role in the production of energy for tumor cells \cite{2016glucose}. However, in this study, only the effects of oxygen concentration on tumor growth are investigated \cite{2003BREWARD,Byrne2002,lewin2020three,2023two}. If the tumor grows with equal oxygen concentrations at tumor boundaries, the tumor grows symmetrically about its center, and the necrotic core also forms symmetrically at the center of the tumor (Fig. \ref{fig:volume}). Also, in this case, the tumor proliferating rim of the same size is observed at both the outer edges. Such qualitative behaviors have been reported in many studies \cite{2017reshmi,PhysRevE.71.051910,ward1997mathematical}. However, in in-vivo situations, it has been seen that the oxygen distribution in the tumor is not uniform. The parts of tissue near the blood vessels have a higher level of oxygen as compared to  that at other parts \cite{2011oxygen_drop}. By considering different oxygen concentrations at boundaries, it is found that the tumor expands more towards the higher level of oxygen as compared to the lower oxygen side (Figs. \ref{fig:volume05}, \ref{fig:volume025}). This eventually leads to an asymmetric shape of tumor as well as a necrotic core is developed. Different widths of the proliferating layers of tumor cells are also obtained at different edges of the tumor. \citet{compton2000prognostic} found that patients with a thicker proliferating layer were more likely to have metastasis. Therefore, the simulation results of the present study may help clinicians to predict the type of tumor or the factors associated with tumor growth.
	
The present model incorporates the stress-free boundary conditions, and the tumor boundary is traced out by tumor cellular  phase velocity. The velocity profile of cellular phase is anti-symmetric about its center (Fig. \ref{fig:velocity}). This kind of profile is achieved when the tumor starts to grow in a medium having equal oxygen concentrations at tumor boundaries. Similar results have also been reported in previous studies \cite{Byrne2002,2023two}. On the other hand, when a tumor grows in a medium where the oxygen concentration available at one boundary is lower than that at the other one, the profile of cellular phase velocity is significantly different from the case of the constant supply of oxygen (Figs. \ref{fig:velocity05}, \ref{fig:velocity025}, \ref{fig:velocity01}). The cellular phase velocity profile is no longer anti-symmetric. If the oxygen level at one boundary is significantly low (i.e., $C_r=0.25\text{ or }0.1$), cellular velocity is nearly zero at that boundary (Figs. \ref{fig:velocity025}, \ref{fig:velocity01}). As a result, tumor growth is negligible at the boundary having a lower level of oxygen supply (Figs. \ref{fig:velocityCompare}). If an ample amount of oxygen is available at tumor boundary, tumor phase velocity increases with time initially and after subsequent times, no change in velocity is observed. This illustrates that the tumor grows exponentially at the initial stage, which is a standard feature and well-established process of tumor growth as reported by the experimental studies \cite{1989exponential,hoffmann2020initial,kiran2009mathematical,2015estimating,yang2020gompertz}. The overall analysis of this article indicates that oxygen concentration at tumor boundary significantly impacts tumor morphology as well as the shape of necrotic core. 

In general, the shape of tumor is spherical, ellipsoidal, or irregular \cite{shape2020}. In the present work,  the model is formulated in one-dimensional cartesian co-ordinate system, the curvature of the general shape is not incorporated. Also, the effects of surface tension is not considered.The present model can be extended to two or three-dimensions, where the effects of surface tension and curvature of the boundary can be incorporated. 
	\section{Conclusions}
	In this study, we have investigated the effects of oxygen concentration on the growth of solid tumors. A one-dimensional (1D) multiphase model is employed to study the growth of tumor which is assumed to be composed of two phases: tumor cellular phase and ECM as the other phase. ECM contains the dead cells as the major component. Therefore, the dynamics of the necrotic core is explored by tracking the dynamics of the ECM phase. The effects of unequal oxygen concentrations on tumor boundaries are explored. It leads to asymmetric growth of the tumor.	Due to the inherent asymmetricity, the pressure, cellular, and ECM velocities are coupled. To solve the model, a numerical method based on the Semi-Implicit Method for Pressure-Linked Equations (SIMPLE) framework is employed. The method uses the staggered idea of finite volume in the finite difference approach to simulate the multiphase model of asymmetric growth.	The dynamics of necrotic core due to unequal oxygen supply at the tumor boundaries and its effects on tumor growth are explored in this study. The following key findings are observed from the simulation results.
	\begin{enumerate}
		\item If the concentration available at one boundary is lower than that at the other boundary, the tumor as well as its necrotic core  grows asymmetrically about the tumor center. Also, the proliferating rims are of different widths at the edge of the tumor.
		
		\item  The necrotic core is closer to the side with lower level of oxygen supply.
		
		\item When a tumor grows with an unequal amount of oxygen at tumor boundaries, the magnitude of cellular phase velocity becomes unequal at the boundaries. Otherwise, it is same for equal oxygen concentrations at the boundaries.
		
		\item If a tumor grows in an environment with unequal oxygen concentrations at boundaries, it takes less time to reach the necrotic threshold value of oxygen concentration in the central region as compared to the case where a tumor grows with equal oxygen concentrations at boundaries. 
		
		\item  It is noticed that a tumor with larger size of necrotic core grows slowly as compared to a tumor containing a smaller size of the necrotic core.
	\end{enumerate}
	 
	The shape of a tumor and position of necrotic core can also have an impact on treatment planning. For example, if a tumor is irregular in shape and has tentacle-like extensions, surgical removal may be more difficult and may necessitate extra therapies such as, chemotherapy or radiation therapy.
	The outcomes of this study can help clinicians optimize chemotherapy and radiation therapy treatments.
	
	\section*{Acknowledgments}
	The first author of this article thanks to the Ministry of Education, Govt. of India, for fellowship and Indian Institute of Technology Guwahati, India for the support provided during the period of this work.
	\section*{Funding}
	The authors did not receive support from any organization for the submitted work. 
	\section*{Declaration of Competing Interest}
	The author declares that there are no conflicts of interest.
	\section*{Ethical approval}
	The authors did not conduct any research with humans or animals.
	\bibliography{mybibfile11}
	\bibliographystyle{elsarticle-num}

\end{document}